\documentclass[fleqn,usenatbib]{mnras}

\pdfoutput=1
\usepackage{mathptmx}
\usepackage[T1]{fontenc}
\usepackage{ae,aecompl}

\usepackage{graphicx}	
\usepackage{amsmath}	
\usepackage{amssymb}	
\usepackage{multirow}

\usepackage{ulem}
\usepackage[T1]{fontenc} 
\usepackage{color}

\newcommand{\df}{{\rm d}}
\newcommand{\mbD}{\pmb{D}}
\newcommand{\mbTh}{\pmb{\Theta}}

\newcommand{\lhood}{\mathcal{L}}
\newcommand{\ev}{\mathcal{Z}}

\newcommand{\mcB}{\mathcal{B}}

\def\be{\begin{equation}}
\def\ee{\end{equation}}
\def\beq{\begin{equation}}
\def\eeq{\end{equation}}
\def\bc{\begin{center}}
\def\ec{\end{center}}
\def\bea{\begin{eqnarray}}
\def\eea{\end{eqnarray}}

\def\MN{{\sc MultiNest}}
\newcommand{\tstat}{\mathcal{T}}

\newcommand{\rsun}{{\rm R_\odot}}

\newcommand{\ysur}{Y_{\rm S}}

\newcommand{\rcz}{R_{\rm CZ}}

\newcommand{\phib}{\Phi({\rm ^8B})}
\newcommand{\phibe}{\Phi({\rm ^7Be})}
\newcommand{\phin}{\Phi({\rm ^{13}N})}
\newcommand{\phio}{\Phi({\rm ^{15}O})}
\newcommand{\phif}{\Phi({\rm ^{17}F})}
\newcommand{\phipp}{\Phi({\rm pp})}
\newcommand{\phipep}{\Phi({\rm pep})}
\newcommand{\phihep}{\Phi({\rm hep})}

\newcommand{\dk}{\delta\kappa}
\newcommand{\lgT}{\log_{10} T}

\title[Data Driven Reconstruction of Solar Properties]{Helioseismic and Neutrino Data Driven Reconstruction of Solar Properties}

\author[N. Song et al.]{Ningqiang Song,$^{1}$\thanks{E-mail: ningqiang.song@stonybrook.edu}
M.C. Gonzalez-Garcia,$^{2,3,1}$ Francesco L. Villante,$^{4,5}$ \newauthor
 Nuria Vinyoles,$^6$ Aldo Serenelli$^6$
\\
$^{1}$C.N.~Yang Institute for Theoretical Physics, State
  University of New York at Stony Brook, Stony Brook, NY 11794-3840,
  USA \\
$^{2}$Departament de Fis\'{\i}ca Qu\`antica i
  Astrof\'{\i}sica and Institut de Ciencies del Cosmos, Universitat de
  Barcelona, Diagonal 647, \\ E-08028 Barcelona, Spain \\
$^{3}$Instituci\'o Catalana de Recerca i Estudis
  Avan\c{c}ats (ICREA), Pg. Lluis Companys 23, 08010 Barcelona,
  Spain \\
$^{4}$Dipartimento di Scienze Fisiche e Chimiche,
  Universit\`a dell'Aquila,   I-67100 L'Aquila, Italy \\
$^{5}$Istituto Nazionale di Fisica Nucleare (INFN) Laboratori Nazionali del Gran Sasso (LNGS),   I-67100 Assergi (AQ), Italy \\
$^{6}$Institute of Space Sciences (IEEC-CSIC), Campus UAB, Carrer de Can Magrans s/n,
E-08193 Bellaterra, Spain
}

\pubyear{2017}

\begin{document}
\label{firstpage}
\pagerange{\pageref{firstpage}--\pageref{lastpage}}
\maketitle

\begin{abstract}
  In this work we use Bayesian inference to quantitatively
  reconstruct the solar properties most relevant to  the solar
  composition problem using as inputs the information provided by
  helioseismic and solar neutrino data. In particular, we use a Gaussian
  process to model the functional shape of the opacity uncertainty to gain
  flexibility and become as free as possible from prejudice in this regard.
  With these tools we first readdress the statistical significance of
the solar composition problem. Furthermore, starting from a
composition unbiased set of standard solar models we are able to
statistically select those with solar chemical composition and other
solar inputs which better describe the helioseismic and neutrino
observations.  In particular, we are able to reconstruct the solar
opacity profile in a data driven fashion, independently of any
reference opacity tables, obtaining a 4\% uncertainty at the base of
the convective envelope and 0.8\% at the solar core. When systematic
uncertainties are included, results are 7.5\% and 2\% respectively.
In addition we find that the values of most of the other inputs of the
standard solar models required to better describe the helioseismic and
neutrino data are in good agreement with those adopted as the standard
priors, with the exception of the astrophysical factor $S_{11}$ and
the microscopic diffusion rates, for which data suggests a 1\% and
30\% reduction respectively. As an output of the study we derive the
corresponding data driven predictions for the solar neutrino fluxes.
  \end{abstract}




\section{Introduction}

Standard Solar Models (SSMs;
\citealp{Bahcall:1987jc,TurckChieze:1988tj, Bahcall:1992hn,
Bahcall:1995bt, Bahcall:2000nu, Bahcall:2004pz,
PenaGaray:2008qe,Serenelli:2011py,Vinyoles:2016djt}) describe the Sun
present day properties as a result of its evolution starting at the
pre-main sequence. A set of observational parameters are taken as
constraints that SSMs calculations have to satisfy by construction.
They include the present surface abundances of heavy elements and
surface luminosity of the Sun, as well as its age, radius and
mass. The modeling relies on some simplifying assumptions such as
hydrostatic equilibrium, spherical symmetry, homogeneous initial
composition, and evolution at constant mass. SSMs have been constantly
refined by including updated experimental results and observations
regarding the values in physical input parameters. Examples include
the values of the nuclear reaction rates and the surface
abundances. There have also been improvements on the accuracy of the
calculation of the constituent quantities like the equation of state
and the radiative opacity, as well as the inclusion of new physical
effects like the diffusion of elements. 

The Sun {\sl burns}, i.e. it generates power through nuclear fusion,
with the basic energy source being the burning of four protons into an
alpha particle, two positrons, and two neutrinos. Being only weakly
interacting, the neutrinos can exit the Sun relatively unaffected.
Thus they give us the opportunity to learn about the solar interior
and test in an almost direct way our understanding of nuclear energy
production in the solar core \citep{Bahcall:1964gx}. With this
objective the original neutrino experiments were designed, their goal
being somewhat diverted by the appearance of the then called ``solar
neutrino problem'' \citep{Bahcall:1968hc, Bahcall:1976zz}.  Thanks to
the increasing experimental accuracy of the measured neutrino flux in
radiochemical experiments Chlorine \citep{Cleveland:1998nv},
Gallex/GNO \citep{Kaether:2010ag} and SAGE
\citep{Abdurashitov:2009tn}, together with the upcoming of the
real-time experiments, Super-Kamiokande
\citep{Hosaka:2005um,Cravens:2008aa,Abe:2010hy,sksol:nu2014}, SNO
\citep{Aharmim:2011vm} and Borexino
\citep{Bellini:2011rx,Bellini:2013lnn,Bellini:2008mr,Bellini:2014uqa},
we have now reached the solution of the problem. It implied the
modification of the Standard Model (of particle physics) with the
addition of neutrino masses and leptonic mixing which imply both
flavour transition of the solar neutrinos from production to detection
\citep{Pontecorvo:1967fh,Gribov:1968kq}, and non-trivial effects in
their flavour evolution when crossing dense regions of matter, the so
called LMA-MSW flavour transitions \citep{Wolfenstein:1977ue,
Mikheev:1986gs}.

Moreover the Sun {\sl beats}. In first approximation it can be
considered to be a non-radial oscillator and the study of its
frequency pattern offers powerful insights as well (for example, see
\citealt{Basu:2007fp}). In particular, the sound speed as a function
of depth can be reconstructed to precision of order 0.1\%.  The
transition from radiative to convective energy transport, or abrupt
changes in the solar thermal structure from ionization, induce
acoustic glitches that can be precisely localized; thus with that
level of precision it is possible to infer the depth of the convective
envelope with 0.2\% accuracy and the surface helium abundance with
1.5\% accuracy.  This is, the solar structure is well constrained and
the Sun can be used as a solid benchmark for stellar evolution and as
a laboratory for fundamental physics (see e.g.
\citealp{Fiorentini:2000gr,Ricci:2002ra,Bottino:2002pd,gondolo:2009,vinyoles:2015,vinyoles:2016b}).

As usual in the history of physics, better experimental information
opens new questions. So in parallel to the increased precision on both
solar neutrino detection and helioseismic solar results, a new puzzle
emerged in the consistency of SSMs \citep{Bahcall:2004yr}.  SSMs built
in the last decade of the last century had notable successes in
predicting the helioseismology related measurements
\citep{Bahcall:1992hn,Bahcall:1995bt,
jcd:1996,Bahcall:2000nu,Bahcall:2004pz}.  A key element to this
agreement was the input value of the abundances of heavy elements on
the surface of the Sun used to compute SSMs
\citep{Grevesse:1998bj}. But in the second half of the first decade of
the 21st century new determinations of these abundances became
available and they pointed towards substantially lower values
\citep{Asplund:2004eu,Asplund:2009fu}. The SSMs built incorporating
such lower metallicities failed at explaining the helioseismic
observations \citep{Bahcall:2004yr}.

So far there has not been a successful solution of this puzzle as no
obvious changes in the Sun modeling have been found which could be
able to account for this discrepancy~
\citep{Castro:2006qu,Guzik:2010ck,Serenelli:2011py}. This has led to
the construction of two different sets of SSMs, one based on the older
solar abundances \citep{Grevesse:1998bj} implying high metallicity,
and one assuming lower metallicity as inferred from the ``newer''
determinations of the solar abundances
\citep{Asplund:2004eu,Asplund:2009fu}.  In a subsequent set of works
\citep{Serenelli:2009yc,Serenelli:2011py,Vinyoles:2016djt} the
neutrino fluxes and helioseismic predictions corresponding to such two
models were detailed, based on updated versions of the solar model
calculations.

Alternatively, attempts to use the information from helioseismic and
neutrino observations to better determine the solar chemical
composition and other solar properties started to be put forward
\citep{Delahaye:2005ed,Villante:2013mba}.  The technical complication
arises from the fact that both neutrino and helioseismic results are
{\sl outputs} of the standard solar model simulations while chemical
composition and the other properties to be inferred are {\sl
inputs}. We are faced then with a common issue in multivariable
analysis, the consistent estimation of the values of input parameters
(some even with unknown functional dependence) which can provide a
valid set of outputs within a given statistical level of agreement
with some data.  Before the advent of fast computing facilities this
could only be attempted by partially reducing the number of inputs to
be allowed to vary. For example, in \citet{Villante:2013mba} the
problem was analyzed in terms of three continuous multiplicative
factors (the abundance of volatiles, that of refractories and that of
Ne) to parametrize the allowed departures of the standard solar model
inputs from the adopted priors of the two model versions. Furthermore,
for the opacity profile -- an input which is not a parameter but a
function -- assumptions about its functional form and allowed range of
functional variation had to be assumed.

In this work we take a step forward in this program by making use of
Bayesian inference methods applied with specific numerical tools such
as \MN\ \citep{Feroz:2007kg,Feroz:2008xx,Feroz:2013hea} which have
been developed precisely to make such inference in large parameter
spaces which may contain multiple modes and pronounced degeneracies.
Furthermore we introduce the use of Gaussian process (GP), a
non-parametric regression method, to reconstruct the opacity profile
and its uncertainty without assuming a specific functional form.

The outline of the paper is as follows. In Sec.~\ref{sec:framework} we
present a brief introduction to the Bayesian parameter inference
methods which we are using in this work.  Section~\ref{sec:opa}
discusses the issues arising in the parametrization of the opacity
profile function for which we first describe the traditional linear
form in Sec.~\ref{sec:linop}. Section~\ref{sec:GPop} presents the
alternative non-parametric GP method to reconstruct the opacity
profile (see also appendix~\ref{sec:GP} where we introduce the main
concepts in GP method for non-parametric functional reconstruction).
In Section~\ref{sec:signif} we first apply this methodology to
readdress the solar composition problem by evaluating a test of
significance of the two B16 SSMs \citep{Vinyoles:2016djt} and using
the two prescriptions of the profiles of the opacity and its
uncertainty (linear and GP). We find, as expected, that allowing for
the most flexible GP form of the opacity uncertainty profile decreases
the evidence against the B16-AGSS09met model but it is still strongly
disfavoured. Section~\ref{sec:results} contains our evaluation of the
optimum solar composition, opacity profile and other solar parameters,
to describe the helioseismic and neutrino data by Bayesian inference
starting with a composition unbiased set of standard solar
models. As an output of the study we derive the corresponding prediction
for the solar neutrino fluxes. Finally in section~\ref{sec:summary} we
summarize our conclusions. Details of the construction of the
Likelihood function used in the analysis of the helioseismic and neutrino
data are summarized in the Appendix~\ref{sec:appendix}.

\section{Statistical Framework}
\label{sec:framework}
Bayesian inference methods provide a consistent approach to the
estimation of a set of parameters $\pmb{\Theta}$ in a model $M$
for the data $\pmb{D}$. Bayes' theorem states that under the assumption
that a model
$M$ is true, complete inference of its parameters is given by the posterior
distribution,
\begin{equation} 
\label{eq:Bayes_params} 
\Pr( \pmb{ \Theta} | \pmb{D},M) = 
\frac{\Pr(\pmb{D}
|\pmb{\Theta},M)\Pr(\pmb{\Theta}|M)}
{\Pr(\pmb{D}|M)}  = \frac{\lhood(\mbTh)\pi(\mbTh)}{\ev},
\end{equation}
where $\mathcal{L}(\pmb{\Theta})\equiv
\Pr(\pmb{D}|\pmb{\Theta}, M)$ is the \textit{likelihood function}.
The prior probability density of the parameters is given by
$\pi(\pmb{\Theta}) \equiv \Pr(\pmb{\Theta}|M)$, and should
always be normalized, i.e., it should integrate to unity.
Conversely the \textit{evidence}, $\ev_i = \Pr(\pmb{D}|M_i)$, is the
likelihood for the model quantifying how well the model describes the data.

From the posterior distribution one can construct reparametrization
invariant Bayesian credible intervals by defining the ``credible level'' of a
value $\eta = \eta_0$ of a subset of parameters simply as the posterior
volume within the likelihood of that value, 
\be {\rm CL}(\eta_0) =
\int_{\lhood(\eta) > \lhood(\eta_0) } \Pr (\eta | \mbD ) \df \eta.
\label{eq:CL}
\ee
This function can be converted to the ``number of $\sigma$'s'' in
the usual manner as
\be 
S = \sqrt{2} {\rm erfc}^{-1}(1-{\rm CL}). 
\label{eq:S}
\ee

Bayesian statistics is mostly suited to make a {\sl relative} statement about
the plausibility of a given model $M_i$ versus another $M_j$ by comparing their
respective posterior probabilities. This is quantified by means of the
Bayes factor
\beq
\mcB_{ij} = \frac{\ev_i}{\ev_j}
\eeq  
which is the ratio of the evidences. 
Jeffrey's scale is often used for the interpretation of the Bayes factors
(see Table \ref{tab:jeff}).
\begin{table}
  \centering
\begin{tabular}{|c@{\hspace{20mm}}|c|}
	\hline
        $\left|\ln(\mcB_{ij})\right|$ & Strength of Evidence
      \\
      \hline
      $< 1.0$ & Inclusive
      \\
      \hline
      1.0--2.5 & Weak to Moderate 
      \\
      \hline
      2.5--5 & Moderate to Strong
      \\
      \hline
      $>5.0$ & Strong to Very Strong/Decisive
      \\
      \hline
\end{tabular}
\label{tab:jeff}
\caption{Jeffrey's scale for interpretation of the Bayes factors}
\end{table}
This gives what the ratio of posterior probabilities for the models would be if
the overall prior probabilities for the two models were
equal. Or in other words it shows by how much the
probability ratio of model $M_i$  to model $M_j$ changes in the light of the
data, and thus can be viewed as a numerical measure of evidence
supplied by the data in favour of one hypothesis over the other.

It is also possible to make an {\sl absolute} test of significance of
a given model $M$ by using the prior predictive distribution,
which is to be understood as a distribution of the possible
observable outputs $\vec{\cal O}$, 
\be
\Pr(\vec {\cal O}) =
\int \Pr(\vec {\cal O} |\pmb{\Theta}) \pi(\pmb{\Theta})
\df^N \pmb{\Theta}
\ee
to determine the probability distribution function for some statistics
$\tstat(\vec {\cal O})$  and compare it with what was actually observed.
This would be done as usual by  calculating the p-value
\be
p = \Pr( \tstat(\vec {\cal O}) \geq \tstat(\vec {\cal O}^{\rm dat}))
\ee

In this work we use
\MN\ \citep{Feroz:2007kg,Feroz:2008xx,Feroz:2013hea}, a Bayesian
inference tool which, given the prior and the likelihood, calculates
the evidence with an uncertainty estimate, and generates posterior
samples from distributions that
may contain multiple modes and
pronounced (curving) degeneracies in high dimensions.

The general procedure which we will follow is to use MC generated sets of
SSMs obtained for different choices of the model input parameters. These
are 20 quantities:
the  Sun luminosity -- $L_{\odot}$-- ,
the  Sun diffusion, the Sun age -- $t_\odot$-- ,
8 Nuclear Rates -- $S_{11}$, $S_{33}$, $S_{34}$, $S_{17}$, $S_{e7}$, $S_{114}$,
$S_{\rm hep}$, $S_{116}$ --, and
9 Element Abundances -- C, N, O, Ne, Mg,  Si, S, Ar, Fe --. Finally, 
one must also input some parametrization of the  Opacity profile and its
uncertainty (more below).

The sets of models are generated according to priors for these inputs which
reflect our knowledge of those (knowledge which is independent  of the data
used in our analysis).
Generically the priors are assumed to be Gaussian distributed. The numerical
values for the mean and standard distributions for the first 20 inputs are
given in \citet{Vinyoles:2016djt}. The assumed priors for the first 11 inputs
are common to all models generated, while for the abundances there are
two different sets of priors corresponding to high-Z and low-Z compositions
leading to the B16-GS98 and B16-AGSS09met model subsets respectively. 

Following the procedure outlined above we confront these models with
the data from helioseismology and neutrino oscillation experiments
described in the Appendix~\ref{sec:appendix}. They amount to an effective
number of 32  data points from helioseismic data plus
a large number of points from the global analysis of neutrino oscillation
data used in \citet{Bergstrom:2016cbh} with which we build the likelihood
function
\begin{equation}
  -2 \ln \mathcal{L}(\pmb{\Theta})= \sum_{i,j} \frac{({\cal
      O}^{\rm mod}_i(\pmb{\Theta})-{\cal O}^{\rm dat}_i)}
  {\sigma^{\rm dat}_i}(\rho^{\rm dat})^{-1}_{ij} \frac{({\cal O}^{\rm
      mod}_j(\pmb{\Theta})-{\cal O}^{\rm dat}_j)} {\sigma^{\rm
      dat}_j}
\end{equation}
where ${\cal O}^{\rm mod}_i(\pmb{\Theta})$ are the model predicted
values for all these  observables obtained by MC generation for a given
set of values of the model inputs $\pmb{\Theta}$.
The correlation matrix $\rho^{\rm dat}_{ij}=\delta_{ij}$ for $i,j=1,32$
but  it is not diagonal for all the other entries which correspond to the
neutrino oscillation data. 
Effectively the neutrino oscillation part of the likelihood can
be approximated by 8 data points corresponding to the extracted solar
flux normalizations given in the last column in table \ref{tab:ssmres}
and with the correlation matrix  in Eq.~\eqref{eq:rhonu}.

With these likelihood functions we can obtain the posterior distribution
for some  (or all) of the input parameters. These posterior distributions
will quantify how the inclusion of this additional data affects our knowledge
of those  properties of the Sun.

Also, as described above, we can use the prior predictive distribution
corresponding to the two variants of the SSM's to carry out a test of
significance and to obtain their corresponding p-values.  For this, we
  define the statistical test $\tstat(\vec {\cal O})$ (where $\vec{\cal O}$ is
an n-dimensional vector containing possible values for the   n
observables):
\begin{equation}
\tstat(\vec {\cal O})
  = (\vec {\cal O} - \vec \langle \vec {\cal O}^{\rm mod}\rangle)^T
  (C_{\rm dat} + C_{\rm mod})^{-1}(\vec {\cal O} - \vec \langle
  \vec {\cal O}^{\rm mod}\rangle) 
\label{eq:T}  
\end{equation}  
where $C_{{\rm dat},ij}=\rho^{\rm  dat}_{ij}\sigma^{\rm dat}_i\sigma^{\rm dat}_j$ is
the covariance matrix associated to the experimental uncertainties, and
\begin{eqnarray}
C_{{\rm mod},ij} & = & 
\langle ({\cal O}^{\rm mod}_i-\bar{\cal O}^{\rm mod}_i)
({\cal O}^{\rm mod}_j -\bar{\cal O}^{\rm mod}_j)\rangle \nonumber \\
 & \equiv & \langle ({\cal O}^{\rm mod}_i-\bar{\cal O}^{\rm mod}_i)\rangle
\rho_{{\rm mod},ij} \langle ({\cal O}^{\rm mod}_i-\bar{\cal O}^{\rm mod}_i)\rangle
\label{eq:rhomod}
\end{eqnarray}
is the model covariance matrix
obtained from the MC generated model
predictions by sampling over the model input priors about their
means $\bar{\cal O}^{\rm mod}_i\equiv\langle {\cal O}^{\rm mod}_i\rangle$. 

The probability distribution of $\tstat(\vec {\cal O})$ can be
determined from the MC model predictions by generating pseudo
experimental results $\vec {\cal O}$ normally distributed according to
$C_{\rm dat}$ around each $\vec {\cal O}^{\rm mod}$ in the MC
generated model samples, and computing for each pseudo experimental
result the corresponding value of $\tstat$. We find that, as expected,
the probability distribution of $\tstat(\vec {\cal O})$ follows very
closely a $\chi^2_{n}$-distribution.

Unlike the first twenty inputs listed above,  the opacity profile is not a
numerical parameter but a function. We describe next two different procedures
to parametrize the uncertainty in its prior.

\section{Treatment of the radiative opacity}
\label{sec:opa}
A fundamentally important physical ingredient in solar models that
cannot be quantified by just one parameter is the radiative opacity,
which is a complicated function of temperature ($T$), density ($\rho$)
and chemical composition of the solar plasma expressed here in terms
of the helium ($Y$) and heavy elements mass fractions ($Z_i$, where
$i$ runs over all metals included in opacity calculations).  In our
calculations, we take as a reference the atomic opacities from OP
\citep{Badnell:2004rz} complemented at low temperatures by molecular
opacities from \citet{Ferguson:2005pu}.  The magnitude and functional
form of its uncertainty is currently not well constrained in available
opacity calculations. As a result, representation of the uncertainty
in radiative opacity by a single parameter \citep{Serenelli:2012zw} or
by taking the difference between two alternative sets of opacity
calculations \citep{Bahcall:2005va,Villante:2013mba} are strong
simplifications, at best. In this paper, instead, we choose to follow
a general and flexible approach based on opacity kernels originally
developed by \citet{Tripathy:1997wy} and later on by
\citet{Villante:2010vt}, which we describe next.

The reference opacity calculation $\bar\kappa(\rho,T, Y, Z_i)$ can be
modified by a generic function of T, $\rho$, $Y$ and $Z_i$. For simplicity,
we assume that opacity variations are parametrized as a function of
$T$ alone such that
\begin{equation}
\kappa(\rho,\,T,\,Y,\,Z_{\rm   i}) = 
\left[1+\delta \kappa_{\rm I} (T) \right] \, \overline{\kappa}(\rho,\,T,\,Y,\,Z_{\rm  i})
\label{prescription} 
\end{equation}
where $\delta \kappa_{\rm I}(T)$ is an arbitrary function  that we call \textit{intrinsic opacity change}. 
The Sun responds linearly even to relatively large opacity variations $\delta \kappa_{\rm I} (T)$  \citet{Tripathy:1997wy,Villante:2010vt}.
Thus, the fractional variation of a generic SSM prediction
\begin{equation}
\delta Q \equiv Q/\bar Q - 1,
\end{equation}
where $Q$ ($\bar Q$) corresponds to the modified (reference) value, can be described as
\begin{equation}
 \delta Q = \int \frac{dT}{T}K_Q (T) \delta \kappa_{\rm I} (T)
 \label{eq:dqkappa}
\end{equation}
by introducing a suitable kernel $K_Q(T)$ that describes the response
of $Q$ to changes in the opacity at a given temperature. We determine
the kernels $K_Q(T)$ numerically by studying the response of solar
models to localized opacity changes as it was done in
\citet{Tripathy:1997wy}. Our results agree very well in all cases
except for variations in the chemical composition because our models
include gravitational settling.

The evaluation of $\delta Q$ is subject to the choice we make for
$\delta \kappa_{\rm I}(T)$. In \citet{Haxton:2009jh} and
\citet{Serenelli:2012zw} the opacity error was modeled as a 2.5\%
constant factor at $1\sigma$ level, comparable to the maximum
difference between the OP and OPAL \citep{Iglesias:1996bh} opacities
in the solar radiative region.  \citet{Villante:2010vt} showed that
this prescription underestimates the contribution of opacity
uncertainty to the sound speed and convective radius error budgets
because the opacity kernels for these quantities are not positive
definite and integrate to zero for $\delta \kappa(T) = {\rm
const}$. Later on, \citet{Villante:2013mba} considered the
temperature-dependent difference between OP and OPAL opacities as
$1\sigma$ opacity uncertainty. However, it is by no means clear that
this difference is a sensible measure of the actual level of
uncertainty in current opacity calculations.

Based on the previous reasons, here we follow a different approach
inspired by the most recent experimental and theoretical results and
some simple assumptions. The contribution of metals to the radiative
opacity is larger at the bottom of the convective envelope ($\sim
70\%$) than at the solar core ($\sim 30\%$).  Moreover, at the base of
the convective envelope, relevant metals like iron are predominantly
in an L-shell configuration, for which atomic models are more
uncertain than for the K-shell configuration that predominates at
solar core conditions.  Also, in a recent theoretical analysis of line
broadening modeling in opacity calculations, \citet{Krief16} have
found that uncertainties linked to it are larger at the base of the
convective envelope than in the core. These arguments suggest that
opacity calculations are more accurate at the solar core than in the
region around the base of the convective envelope. It is thus natural
to consider error parameterizations that allow opacity to fluctuate by
a larger amount in the external radiative region than in the center of
the Sun.

\subsection{Linear Parametrization of Intrinsic Opacity Profile
Uncertainty}
\label{sec:linop} Taking all this into account, we consider the
following parameterization for the intrinsic opacity change (relative
to some reference value): 
\begin{equation} \delta\kappa_I(T) = a + b \; \frac{\log_{10}(T_{\rm
C}/T)} {\tau}
\label{eq:dkilin}
\end{equation} where $\tau = \log_{10}(T_{\rm C}/T_{\rm CZ}) = 0.9$,
$T_{\rm C}$ and $T_{\rm CZ}$ are the temperatures at the solar center
and at the bottom of the convective zone respectively.
This equation is applied only up to the lower regions of the
convective envelope, where convection is adiabatic and changes in the
opacity do not modify the solar structure. Opacity changes in the
uppermost part of the convective envelope and atmosphere are absorbed
in the solar calibration by changes in the mixing length parameter
and, in sound speed inversions, by the surface term. In the context of
SSMs, they will not produce changes in the solar properties considered
in the present work. 

\begin{figure*}
  \centering
  \includegraphics[width=0.8\textwidth]{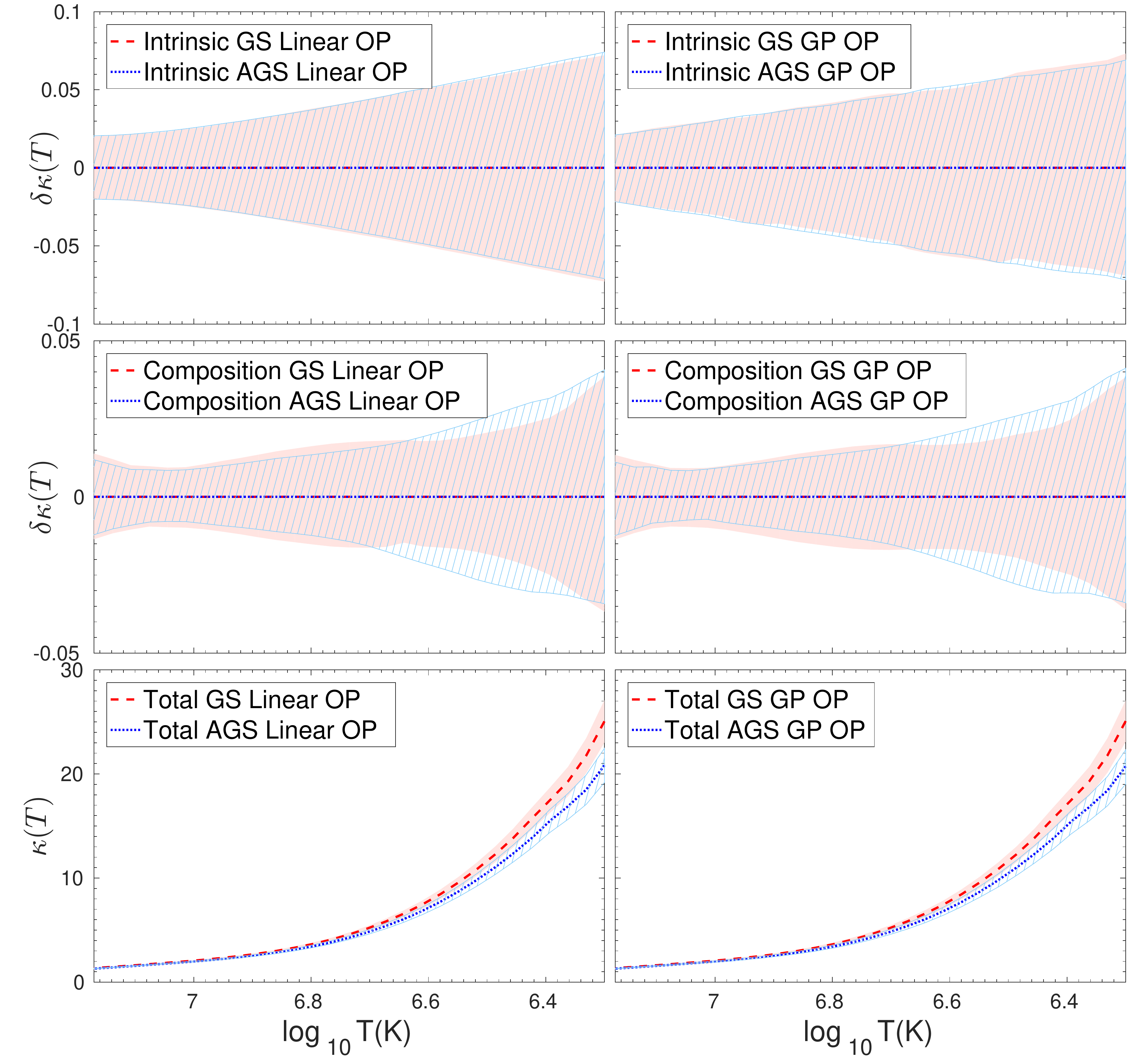}
  \caption{
\label{fig:opaprior}   
    Priors for the intrinsic opacity change (upper panels),
    the composition opacity change, Eq.~\eqref{deltakappaZ} (central panels),
    and the total opacity profile (lower panels). Left correspond to
    the linear parametrization of the intrinsic opacity uncertainty while
    right corresponds to the Gaussian Process one.}
\end{figure*}

Technically, the opacity uncertainty is incorporated in our model
generation by extending the parameter space with 2 more independent
inputs, $a$ and $b$, each with a gaussian prior with zero mean and
variances $\sigma_a$ and $\sigma_b$, respectively. This corresponds to
assuming that the opacity error at the solar center is $\sigma_{\rm
min} = \sigma_a$, while it is given by $\sigma_{\rm out} \simeq
\sqrt{\sigma^2_a + \sigma^2_b}$ at the base of the convective zone. We
fix $\sigma_{\rm in} =\sigma_a = 2\%$ which is the average difference
of the OP and OPAL opacity tables. This is also comparable to
differences found with respect to the new OPAS opacity tables
\citep{mondet:2015} for the AGSS09 solar composition, the only one
available in OPAS.  The more recent OPLIB tables from Los Alamos
\citep{colgan:2016} show much larger differences in the solar core,
about 10 to 12\% lower than OP and up to 15\% lower than OPAS. 
However, OPLIB opacities lead to solar models that predict too low
$Y_S$ and $\phibe$ and $\phib$ fluxes that cannot be reconciled with
data.  For $\sigma_{\rm out}$ we choose 7\%
(i.e. $\sigma_b=6.7\%$), motivated by the recent experimental results
of  \citet{Bailey15} that have measured the iron opacity at conditions
similar to those at the base of the solar convective envelope and have
found a $7\% \pm 4 \%$ increase with respect to the theoretical
expectations. The resulting prior for the intrinsic opacity
profile uncertainty is shown in the upper left panel in
Fig.~\ref{fig:opaprior} for both B16 models.  Given the generated
values for those two parameters we construct the function $\dk_I(T)$
as in Eq.~\eqref{eq:dkilin} and with that we compute the corresponding
change in the output quantities as in Eq.~\eqref{eq:dqkappa}.

As we will see below and was also discussed by \citet{Vinyoles:2016djt},
it turns out that our ad-hoc linear parametrization of the intrinsic
opacity uncertainty is not flexible enough to accommodate the tension
between B16-AGSS09met model and data (especially sound speed
data). This parametrization was chosen for its simplicity, whereas in
fact the shape of the opacity uncertainty function is unknown.  Thus,
in the next section we turn to a more general modeling of the 
intrinsic opacity uncertainty
based on a Gaussian Process approach. A brief introduction of the general 
method is given in Appendix~\ref{sec:GP}.
 
\subsection{Gaussian Process Reconstruction of the 
Opacity Profile}
\label{sec:GPop}
 
Our goal is to define the uncertainty of the opacity profile without
using parameterized functions and to reconstruct the intrinsic opacity
change $\dk_{\rm I}(T)$ that can lead to a better agreement with the
data.
In order to do this, following the discussion in Sec.~\ref{sec:linop},
we assume that $\dk_{\rm I}(T)$ is a gaussian variable with mean is
$\mu_P(T)\equiv 0$ with a temperature dependent variance $\sigma(T)$
which allows for 2\% uncertainty in the solar center and 7\% at the
base of the convective zone, i.e.:
\begin{equation}
  \sigma(T)=0.02+(0.07-0.02)\cdot\frac{\log_{10} (T_{\rm C}/T)}{\tau}\; .
  \label{eq:p1}
\end{equation}  
As the values of the opacity at two different temperatures $T$ and
$T'$ may be not independent, we introduce a prior covariance function
$C_P(T,T')$.  A possible choice is
\begin{equation}
C_P(T,T')=\sigma(T)\sigma(T')\rho(T,T'),
\label{eq:cpop}
\end{equation}
with
\begin{equation}
\rho(T,T')=\mathrm{exp}\left[-\frac{1}{2}\left(\frac{\lgT-\lgT'}{\tau \, L}\right)^2\right]. \label{eq:p3}
\end{equation}
Here $L$ determines the characteristic correlation length over which $\dk_{\rm I}(T)$ can vary significantly and
it is the only hyperparameter in our
analysis. According to the above definition,  $L=1$ means 
maximum correlation between the opacity at the edge of the convective
zone and at the center. 
If $L$ is too large
the correlation is too strong and the model is over constrained.
If, on the other hand, $L$ is too small, we are allowing opacity errors to
dominate the output of solar models, and we can barely learn anything
from the data. Moreover, there is a physically motivated lower bound
for $L$ given by the temperature range over which the opacity can vary
substantially.  In the solar interior $|\partial \ln{\kappa} /
\partial \ln{T}| < 2$  \citep{colgan:2016}.  From this, the smallest
temperature range over which $\Delta \ln{\kappa} \approx 1$ is $\Delta
\ln{T} \approx 0.5$, i.e. $L_{\rm min} \approx 0.2$.

To implement the Gaussian Process intrinsic opacity uncertainty in our
analysis we start by choosing a set of $N$ points in which we evaluate
the function $\dk_{{\rm I} i}=\dk_{\rm I}(T_i)$. We take $N=11$ and
choose the points to be uniformly distributed in $\lgT$ between 6.3
and 7.2.  The parameter space for model generation is thus extended
with 12 more input parameters: the length $L$ and the 11 $\dk_{{\rm I}
i}$ values.  For the correlation length we assume a uniform prior in
$\log_{10} L$ between $\log_{10}0.2$ and $\log_{10}1$.  The values
$\dk_{{\rm I} i}$ are generated according to prior distribution
defined by Eqs.\,(\ref{eq:p1}-\ref{eq:p3}), together with 
$\mu_P(T)\equiv0$\footnote{We
  studied the number of points which maximized the smoothness of the
  output profile versus computing time and found that increasing $N$
  beyond 11 did not yield any better results.}.  
Given a set of values for the 11 $\dk_{Ii}$'s we construct the full
function $\dk_I(T)$ by linear interpolation between these values and
with that we compute the corresponding change in the output
predictions as in Eq.~\eqref{eq:dqkappa}.  The resulting prior
distribution of the intrinsic opacity change is shown in the upper
right panel of Fig.~\ref{fig:opaprior}.  As seen in the figure, the
ranges of uncertainty profiles for the linear parametrization and the
GP opacity are very similar.  They do, however, lead to different
conclusions when testing the SSM's models versus the data (in
particular vs helioseismic data) as described in
Sec.~\ref{sec:signif}.

\subsection{The Degeneracy between Opacity and 
Composition Effects}
\label{sec:opavscomp}

The properties of the Sun depend on its \textit{opacity profile}
$\kappa_{\rm SSM}(T)$ that we define as:
\begin{equation}
\kappa_{\rm SSM}(T) =  \kappa(\rho(T), \, T, \,Y(T),\, Z_{i}(T))
\label{kappassm}
 \end{equation}
where $\rho(T)$, $Y(T)$ and $Z_{i}(T)$ describe the density, helium
and heavy element abundances stratifications as a function of the
temperature of the solar plasma in a given SSM.
This is indeed the quantity that determines the efficiency of
radiative energy transport and, thus, the temperature gradient at each
point of the Sun and that is plotted in the lower panels of
Fig.~\ref{fig:opaprior} for both B16 models.
When considering an intrinsic opacity change $\delta \kappa_{\rm
I}(T)$ and/or other input parameters in the SSMs are varied, the SSM
needs to be recalibrated, thus obtaining different density and
chemical abundances stratifications with respect to the reference SSM.
As a consequence, the total variation of the solar opacity profile is given by: 
\begin{eqnarray}
 \nonumber
 \delta \kappa_{\rm SSM}(T) &\equiv& \kappa_{\rm   SSM}(T)/\overline{\kappa}_{\rm SSM}(T) - 1 \simeq\\
 &\simeq & 
  \delta \kappa_{\rm I}(T) + \frac{\partial \ln \kappa}{\partial \ln \rho} \delta \rho(T) 
  + \frac{\partial \ln \kappa}{\partial \ln Y} \delta Y(T)  \nonumber
 +  \\ && + \sum_i \frac{\partial \ln \kappa}{\partial \ln Z_i} \delta Z_i(T) 
\label{deltakappassm}
 \end{eqnarray}
where $\delta \rho(T)$, $\delta Y(T)$ and $\delta Z_i(T)$ are the
fractional variations of density and elemental abundances in the
perturbed Sun with respect to the reference SSM, evaluated at a fixed
temperature $T$.

As discussed above, the metal abundances $Z_i(T)$ are derived
quantities that have to be obtained as a results of numerical solar
modeling. However, when we consider a modification of the surface
composition $\{z_i\}$, expressed here in terms of the quantities
$z_{i}\equiv Z_{i,\rm S} / X_{\rm S}$ where $Z_{i,\rm S}$ is the
surface abundance of the $i-$element and $X_{\rm S}$ is that of
hydrogen, we can approximately assume $\delta Z_{i}(T) \simeq \delta
z_{i}$ where $\delta z_{i}$ is the fractional variation of $z_{i}$
with respect to some reference value. 
As a consequence, Eq.~\eqref{deltakappassm} 
can be rewritten as: 
\begin{equation}
 \delta \kappa_{\rm SSM}(T) = 
  \delta \kappa_{\rm I}(T) + 
 \delta \kappa_{\rm Z}(T) +
\frac{\partial \ln \kappa}{\partial \ln \rho} \delta \rho(T) 
  + \frac{\partial \ln \kappa}{\partial \ln Y} \delta Y(T) 
 \end{equation}
where the \textit{composition opacity change} $\delta \kappa_{\rm Z}(T) $ is defined as:
\begin{equation}
 \delta \kappa_{\rm Z}(T) \equiv
\sum_i \frac{\partial \ln \kappa}{\partial \ln Z_i} \delta z_i
\label{deltakappaZ}
\end{equation}
We define the total opacity change $\delta \kappa(r)$ as:
\begin{equation}
\delta \kappa(T) =  \delta \kappa_{\rm I}(T)  + \delta \kappa_{\rm Z}(T),
\label{deltakappa}
\end{equation}
which groups together the contributions to $\delta \kappa_{\rm
SSM}(T)$ directly related to the variations of the input parameters.

Note that metals have a negligible role in determining the equation of
state of the solar plasma and in solar energy generation (except for
carbon, nitrogen and oxygen that determine the efficiency of the CNO
cycle which is, however, largely subdominant in the Sun). Thus, the
only structural effect produced by a modification of the surface
composition $\{z_i\}$ is through the changes in the efficiency of
radiative energy transport induced by the composition opacity change
$\delta \kappa_{\rm Z}(T)$ defined above.  In this respect,
Eq.~\eqref{deltakappa} although being approximate, is quite useful
because it makes explicit the connection (and the degeneracy) between
the effects produced by an intrinsic modification of the radiative
opacity and those produced by a modification of the heavy element
admixture. The physical quantity that drives the modification of the
solar properties and that can be constrained by observational data
is the total opacity change $\delta \kappa(T)$, not the intrinsic $\delta
\kappa_{\rm I}(T)$ or the composition opacity change $\delta
\kappa_{\rm Z}(T)$ separately.

For completeness, we show in the middle panels of
Fig.~\ref{fig:opaprior} the prior distributions of the composition
opacity changes for both B16-SSMs, calculated by considering the
relative variations of the individual abundances $\delta z_j$ around
their mean values for GS98 and AGSS09met surface compositions.  The
logarithmic derivatives $\partial \ln \kappa / \partial \ln Z_i$ can
be found in the left panel of Figure~10 in
\citet{Villante:2013mba}. The prior distributions for $\delta
\kappa_{\rm Z}(T)$ are identical in the left and right (middle) panels
because the adopted procedure for describing the intrinsic opacity
uncertainty does not alter the sampling in surface composition.

\section{Test of Significance and Model Comparison}
\label{sec:signif}

We start by performing a test of significance of the two B16 SSMs
using the linear and GP models of the opacity uncertainty described in
the previous section.  Results are given in Tab.\,\ref{tab:T} where we
show the value of the test statistics $\tstat$ in Eq.~\eqref{eq:T} 
for different combination of observables.
\begin{table*}
\centering
\setlength{\tabcolsep}{2.5pt}
\begin{tabular}{c|c|cccc|cccc}
  \multicolumn{2}{c}{} &
    \multicolumn{4}{c}{LIN-OP} &
    \multicolumn{4}{c}{GP-OP} \\\hline
    \multicolumn{2}{c}{} &
  \multicolumn{2}{c}{GS98 } & \multicolumn{2}{c}{AGSS09met }
  &\multicolumn{2}{c}{GS98 } & \multicolumn{2}{c}{AGSS09met }
  \\ \hline
  $\vec {\cal O}$ & \multicolumn{1}{|c|}{n}
  &  $\tstat(\vec {\cal O})$
   & p-value\,$(\sigma)$ & $\tstat(\vec{\cal O})$ & p-value\,$(\sigma)$
&  $\tstat(\vec {\cal O})$
   & p-value\,$(\sigma)$ & $\tstat(\vec{\cal O})$ & p-value\,$(\sigma)$
  \\ \hline
  $\ysur + \rcz$  & \multicolumn{1}{|c|}{2} & 0.9 & 0.5 & 6.5 & 2.1
&0.7 & 0.35 & 6.9 & 2.2 \\
$\delta c$   & \multicolumn{1}{|c|}{30} & 58.0 & 3.2 & 76.1 & 4.5
& 35.6 & 1.2 & 40.2 & 1.6 \\  
  all $\nu$-fluxes  & \multicolumn{1}{|c|}{8} & 6.0 & 0.5 & 7.0 & 0.6
& 5.9 & 0.44 & 7.0 & 0.6 \\  
\hline
global & \multicolumn{1}{|c|}{40} & 65.0 & 2.7 & 94.2 & 4.7
& 45.1 & 1.1 & 57.1 & 2.1 \\ \hline
\end{tabular}
\caption{Comparison of B16 SSMs against different ensembles of solar
  observables. \label{tab:T}}
\end{table*}
As seen from the table, global p-values are dominated by the sound speed
for both models, although $\ysur$ and $\rcz$ are also relevant for
B16-AGSS09met. We also read  from the table that  when using the linear
opacity uncertainty parametrization the global analysis yields a not too good
p-value of 2.7$\sigma$  for B16-GS98 and considerably worse (4.7$\sigma$) for
the B16-AGSS09met. The results are different when the
GP opacity uncertainty is used  which yields p-value of 1.1$\sigma$
and 2.1 $\sigma$ for B16-GS98 and B16-AGSS09met, respectively.  

\begin{figure*}\centering
\hspace*{0.6cm}\includegraphics[width=0.72\textwidth]{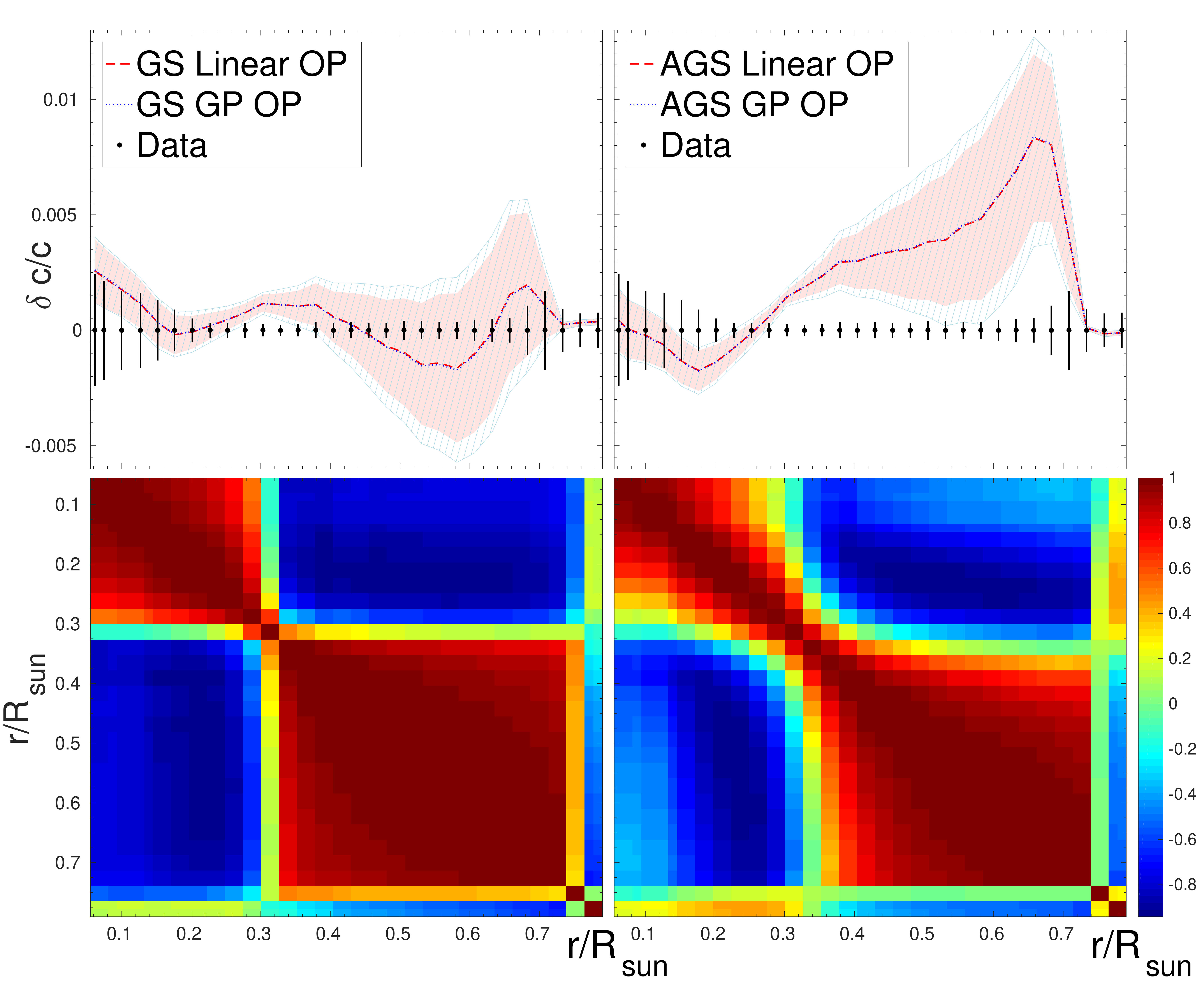}
\caption{
\label{fig:cspriors}%
  $1\sigma$ range of variation of the fractional sound speed profiles
  as predicted by the priors the B16 SSM models and for both opacity profile
  priors discussed in the
  text (upper panels) compared with the 30 data points used in the analysis.
  The lower panels graphically display the values of the  entries in
  the $30\times 30$ model correlation matrix between between the predicted
  sound speeds at the 30  points (which are the same for
  B16-GS98  and B16-AGSS09met models) for the linear opacity uncertainty
  parametrization
  (left) and the GP opacity uncertainty(right).}
\end{figure*}

In Fig.~\ref{fig:cspriors} we plot the fractional sound speed 
difference $\delta c(r) \equiv (c_{\rm obs}(r) -c(r))/c(r)$, where 
$c_{\rm obs}(r)$ is the sound speed inferred from helioseismic data 
while $c(r)$ represent the sound speed profile predicted by the B16-GS98 (left) 
and B16-AGSS09met (right) model, respectively.
The blue (lighter) hatched area and the red (darker) shaded area
corresponds to the 
1$\sigma$ theoretical uncertainties in sound speed predictions obtained
for linear and GP opacity uncertainty priors.
As seen from the figure, and expected from the comparison of the top panels in 
Fig.~\ref{fig:opaprior}, 
they are not very different in the two considered cases.  
Moreover,  we observe from the figure that at almost all radii,
independently of the adopted prescription,
the sound speed profile of B16-GS98 fits well within the 1$\sigma$
data uncertainties. 
It may be thus surprising that the B16-GS98 model
is not providing a good p-value in the case of linear opacity uncertainty
parametrization.

The reason for the bad p-value obtained for the B16-GS98  model 
is that, as discussed in
\citet{Vinyoles:2016djt}, changes in input quantities do not lead to
variations in SSM sound speeds on very small radial scales, so values
of the sound speed at different radii in solar models are strongly
correlated, i.e.  the model correlation matrix $\rho_{{\rm mod},ij}$
in Eq.~\eqref{eq:rhomod} is highly non-diagonal.
This is shown in the lower panels of Fig.~\ref{fig:cspriors} where we
graphically display the values for the entries of the correlation
matrix between the predicted sound speeds at the 30 locations (the
correlation matrix is the same for both B16 models).
As seen in the figure, the characteristic correlation length (i.e. the
distance $|r_i-r_j|$ over which correlations between the predicted
values of the sound speeds are strong, say $|\rho_{{\rm mod}, ij}|
\gtrsim 0.5$) is much larger for the linear opacity profile 
parametrization than for the GP profile.

\begin{figure}
  \centering
  \includegraphics[width=0.4\textwidth]{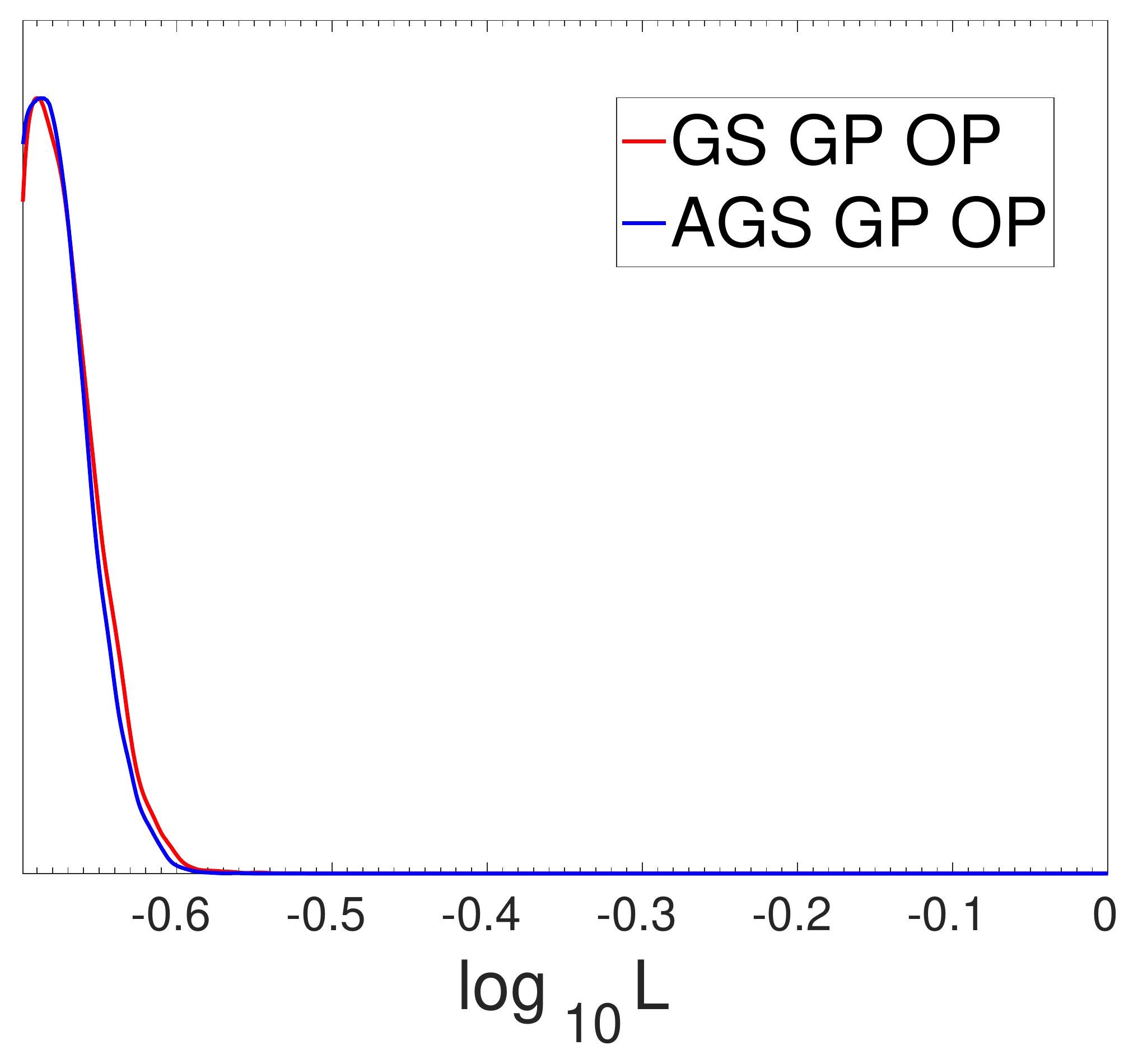}
  \caption{
  \label{fig:Lmodpost}
    Posterior probability distribution for the correlation length
    hyperparameter $L$ of the GP opacity uncertainty for runs with the
    models B16-GS98 and B16-AGSS09met priors for the abundances.}
\end{figure}

The more flexible implementation of the opacity profile uncertainty
provided by the GP procedure permits to obtain a better description of
the observational data for both B16-GS98 and B16-AGSS09met models.
To illustrate this point, Fig.~\ref{fig:Lmodpost} shows the posterior
distribution of $L$, the correlation length hyperparameter
(Eq.~\eqref{eq:cpop}).
As seen from the figure, the best possible description of the data
is achieved with correlation lengths of average $\langle L\rangle\sim 0.2$, i.e. 
close to the lowest value permitted by the adopted prior that allows for short scale 
modifications of the sound speed profiles.

We finish this section by giving in Table~\ref{tab:bayesmod} the Bayes
factors for the two models as obtained with their posterior
probability distributions after including the neutrino and
helioseismic data for the two assumed opacity profile uncertainties.
From the table we conclude that the B16-AGSS09met models are always
somewhat disfavoured with respect to the B16-GS98 model by all data
sets but the most statistical significant effect is driven by the
sound speed profile data. This is particularly the case for the linear
opacity uncertainty profile for which the Bayes factor of -14.7 is
enough for rejection of the model.  Allowing for the most flexible GP
form of the opacity uncertainty
decreases the evidence against the B16-AGSS09met model to close
to strong.

\begin{table}
  \centering
\begin{tabular}{|l||c|c|} \hline
  & \multicolumn{2}{c|}{B16-AGSS09met/B16-GS98}\\ \hline
  Data & LIN-OP & GP-OP\\ \hline
 $\nu$  & -0.23 & -0.27 \\ 
+$Y_S$+$R_{CZ}$ & -1.6 & -2.2 \\ 
+ sound speeds &   -14.7 & -4.1 \\ \hline 
\end{tabular}
\caption{Bayes factor, $\ln(\mcB)$,
  for the B16-AGSS09met  vs B16-GS98 model obtained
  with the  different data sets (see table \ref{tab:jeff} for interpretation).
  \label{tab:bayesmod}}
\end{table}

\section{Determination of the Optimum Composition and  Opacity Profile}
\label{sec:results}

We now turn to the determination of the optimum solar composition which  
best describes the helioseismic and neutrino data.
In order to do so we perform Bayesian parameter inference by using a top
 hat prior for the logarithmic abundances $\varepsilon_i \equiv
 \log_{10}(N_i /N_{\rm H}) + 12$ that accommodates both the
 AGSS09met and GS98 admixtures, i.e. with value 1 between the 
 $3\sigma$ lower value of the AGSS09met composition
and the   $3\sigma$ upper value of the GS998 composition 
for all the nine elements relevant for solar model construction given in
Tab.~\ref{tab:compost}, and zero outside this range.
As before we study the dependence of our results on the two models for 
the opacity uncertainty. 
\begin{figure*}
\centering
\includegraphics[width=0.6\textwidth]{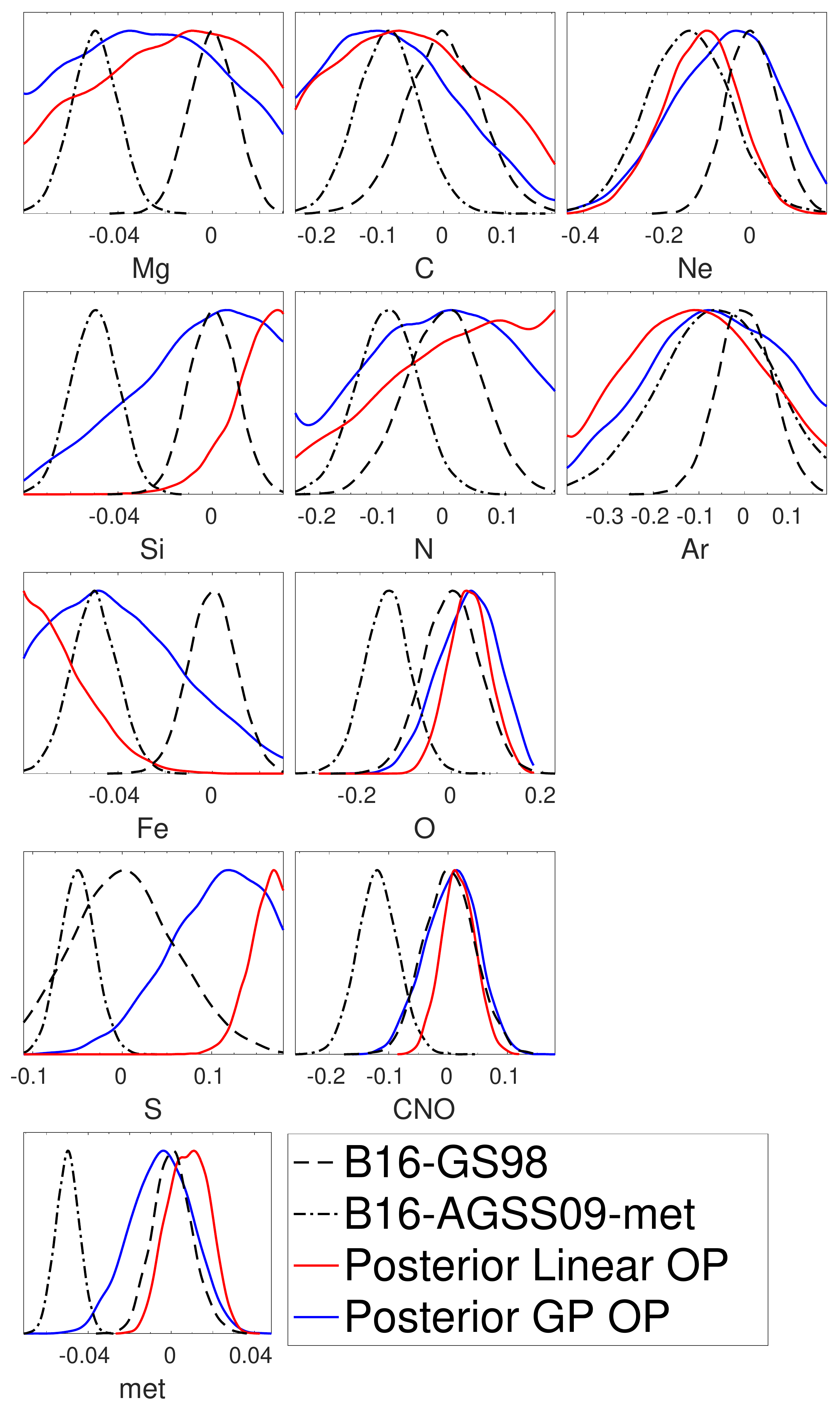}
\caption{
 \label{fig:comppostUNI}
   Posterior probability distribution for the logarithmic abundances
   ($\varepsilon_j-\varepsilon_{j,\rm GS98})$ from the analysis of neutrino and 
   helioseismic
   data  with uniform priors for the abundances  and for the two choices of
   the prior opacity uncertainties. The distributions are given in
   arbitrary units and they have been normalized in such a way that the
   maximum of all distributions lays at the same height.
   See text for details.}
\end{figure*}

\begin{figure}
\centering
\includegraphics[width=0.45\textwidth]{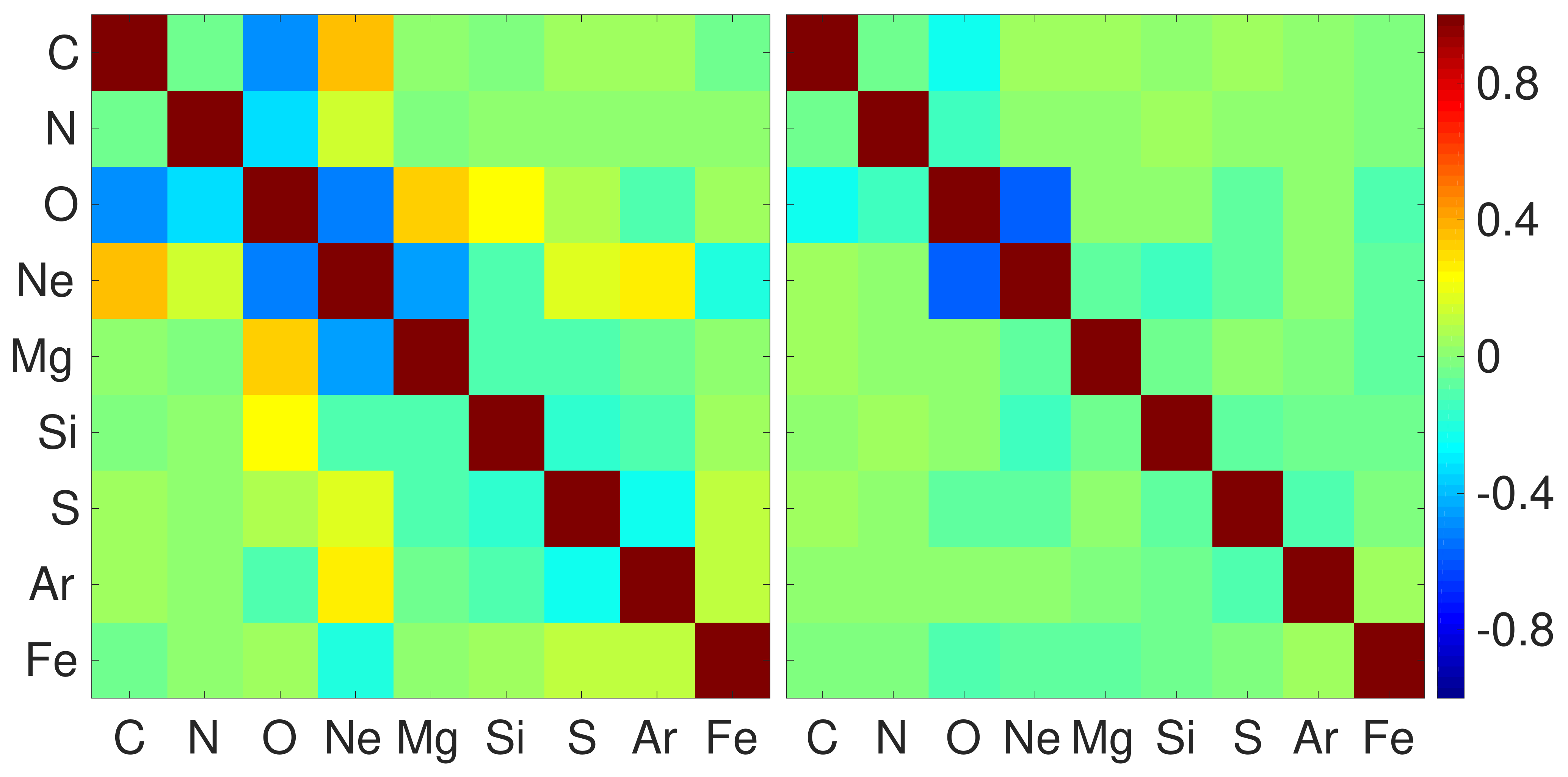}
\caption{Graphical representation of correlations between the
posterior distributions of individual elemental abundances (elements
in the same order as in Table~\ref{tab:compost}) for the linear (left)
and GP (right) models of intrinsic opacity
uncertainty. 
\label{fig:comppostcorrUNI}}
\end{figure}  
    
\setlength{\tabcolsep}{4pt}
\begin{table}
\centering
\begin{tabular}{c| c c c c}
  Element &  GS98 & AGSS09met
  & Linear & GP \\
\hline
 C & $8.52  \pm 0.06$ & $8.43 \pm 0.05$ & [8.32, 8.56] & [8.31, 8.51] \\
 N & $7.92  \pm 0.06$ & $7.83 \pm 0.05$ & [7.88, 8.10] & [7.81, 8.05] \\
 O & $8.83  \pm 0.06$ & $8.69 \pm 0.05$ & [8.82, 8.91] & [8.80, 8.94] \\ 
 Ne & $8.08 \pm 0.06$ & $7.93 \pm 0.10$ & [7.87, 8.06] & [7.90, 8.16] \\
 Mg & $7.58 \pm 0.01$ & $7.53 \pm 0.01$ & [7.54, 7.60] & [7.52, 7.58] \\
 Si & $7.56 \pm 0.01$ & $7.51 \pm 0.01$ & [7.57, 7.59] & [7.54, 7.59] \\
 S  & $7.20 \pm 0.06$ & $7.15 \pm 0.02$ & [7.35, 7.38] & [7.27, 7.37] \\
 Ar & $6.40 \pm 0.06$ & $6.40 \pm 0.13$ & [6.14, 6.44] & [6.20, 6.50] \\
 Fe & $7.50 \pm 0.01$ & $7.45 \pm 0.01$ & [7.42, 7.44] & [7.42, 7.48] \\
\hline
CNO & $9.04\pm 0.04$     &  $8.92\pm 0.03$ &  [9.03, 9.08] & [9.00, 9.09]\\
meteor.& $8.09\pm 0.01$ &  $8.04\pm 0.01$ & [8.08, 8.10] &[8.07, 8.10]\\
\hline
\end{tabular}
\caption{1-$\sigma$ ranges for the logarithmic abundances
$\varepsilon_j$. The first two columns show the mean values and
uncertainties of the GS98 and AGSS09met heavy element admixtures.  The
last two columns give the ranges of the posterior distributions from
the analysis of neutrino and helioseismic data for the two choices of
the prior opacity uncertainties with uniform priors for the
abundances.
\label{tab:compost}}
\end{table}

We show in Fig.~\ref{fig:comppostUNI} the posterior probability
distributions for the nine abundance parameters centered for reference
around the GS98 ones, i.e. $\Delta \varepsilon_j=\varepsilon_j -
\varepsilon_{j,\rm GS98}$, and for the two choices of priors of the
opacity uncertainties (Linear or GP).  The window for each abundance
corresponds to the allowed range, i.e. where prior=1.  Outside each
window the value of the prior is zero.  For the sake of comparison we
also show in the figure the corresponding prior distributions for the
B16-GS98 and B16-AGSS09met models. Notice that the distributions are
given in arbitrary units and have been normalized in such a way that
the maximum of all distributions lays at the same height to help
comparison.

We list in the last two columns of Table~\ref{tab:compost} the
corresponding $\pm 1$-$\sigma$ ranges for the logarithmic abundances
$\varepsilon_j$ extracted from these posterior distribution. These can
be compared with the determination of the same quantities in GS98 and
AGSS09met compilations reported in the first two columns of the same
table.  From the figure and table we see that the available data are
not capable of setting tight constraints on all the elements
simultaneously.  However we find that the posterior for the
combinations of CNO (C+N+O) and meteorite (Mg+Si+S+Fe) abundances
\citep{Delahaye:2005ed,Villante:2013mba} have a comparable precision
to GS98 and AGSS09met observational determinations for either choice
of the opacity uncertainty parametrization.  It is important to stress
here that the distributions for these combinations have been obtained
without assuming any prior correlation between the individual
elements.  This is in contrast to previous work
\citep{Delahaye:2005ed,Villante:2013mba}, where abundances of all
elements within a group were forced to have the same proportional
change.  Correlations among the posterior distributions of the
abundances appear exclusively as output of the data analysis.  For the
sake of illustration we provide in Fig.~\ref{fig:comppostcorrUNI} a
graphic representation of the correlation among the posterior
probability distributions of the different elemental abundances.
As expected, the correlations are smaller for the run with the more
flexible GP description of the opacity profile uncertainty.  But in
general for both GP and Linear opacity uncertainties, the correlation
among the posterior distributions of the abundances included either
the CNO or the meteorite groups are not very large.
The exception is provided by the large anticorrelation between the  posterior
distributions of C and O for the analysis with Linear opacity uncertainty. 
We have verified that because the allowed ranges of C and O are strongly
anticorrelated in this case,  the allowed range of CNO group abundance
results to be more precise than any of the model priors
as seen in the lower central panel in Fig.~\ref{fig:comppostUNI}.

\begin{figure*}
\centering
\includegraphics[width=0.7\textwidth]{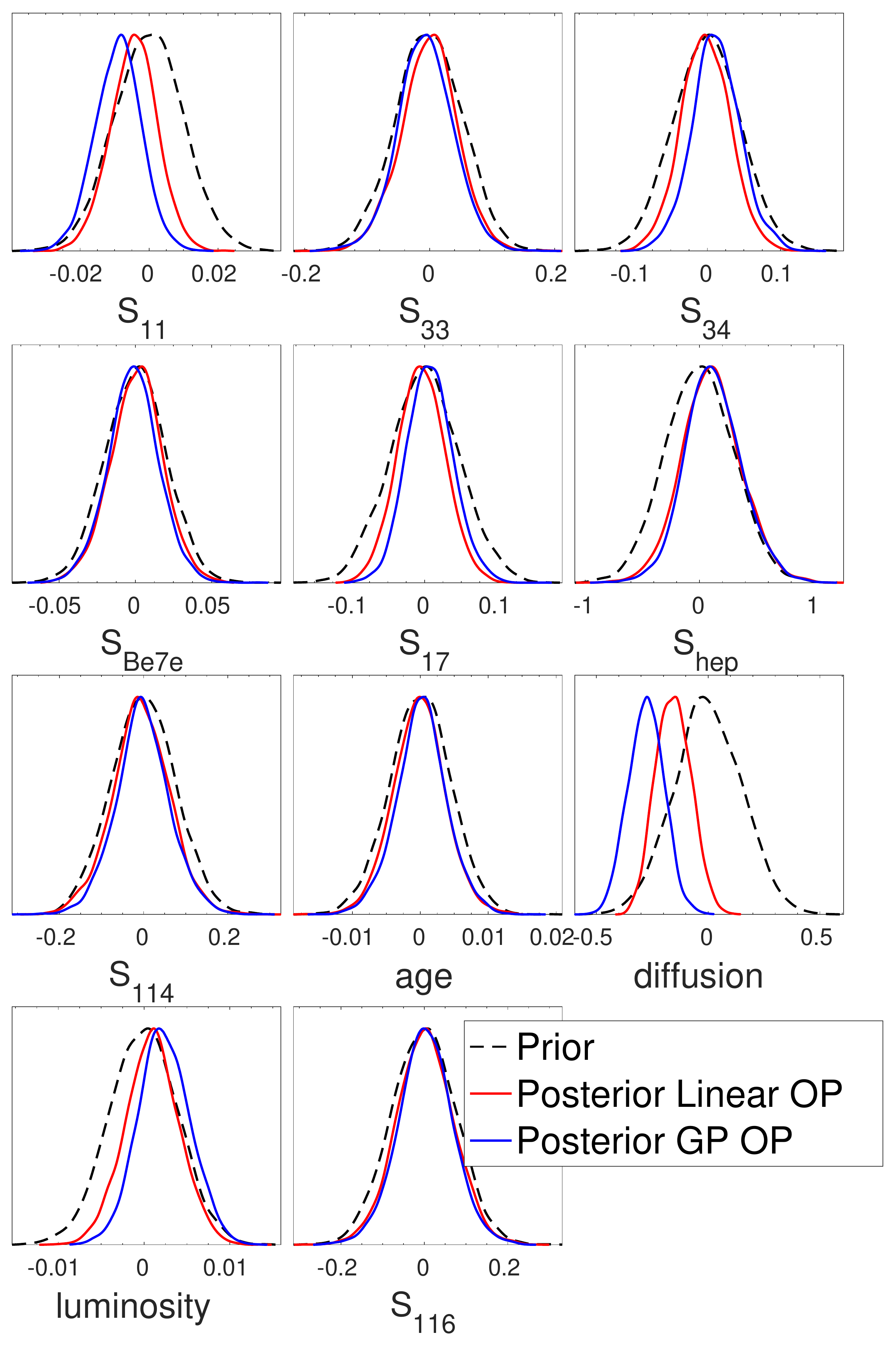}
\caption{\label{fig:postotherUNI}
  Posterior probability distribution for the Sun luminosity, Sun diffusion,
  the Sun age, and  the eight nuclear rates from the analysis of neutrino
  and helioseismic  data  with uniform priors for the abundances  and for
  the two choices of  the prior opacity uncertainties. For comparison we also
  show their prior distribution. The distributions are given in
   arbitrary units and they have been normalized in such a way that the
   maximum of all distributions lay at the same height.} 
\end{figure*}

The posterior distributions for the other solar input parameters --
luminosity, diffusion, age, and the eight nuclear rates,
are shown in Fig.~\ref{fig:postotherUNI} together with their gaussian
priors.  From the figure we see that with the exception of $S_{11}$
and diffusion coefficients, all others parameters do not get
significantly modified with respect to the model priors by the
inclusion of the neutrino and helioseismic data, irrespective of the
form of the opacity uncertainty.  We have verified that the
helioseismic data -- the surface helium abundance $\ysur$ and the
location of the bottom of the convective envelope $\rcz$ -- are the
most relevant in driving the shift in $S_{11}$.  We see from the
figure that the posterior distributions for $S_{11}$ show a preferred
value about 1\% lower than our prior central value taken from
\citep{Marcucci:2013tda} and 1.5\% lower than the newer determination
of $S_{11}$ by \citep{Acharya:2016kfl}.  A reassessment of this
relevant rate might be therefore important.  The sound speed data are
instead responsible for the preference of lower values of the
diffusion coefficients.  The reduction in diffusion efficiency that we
obtain is in line with previous work \citep{Villante:2013mba}. Our
analysis, however, points towards a $30 \pm 10$\% reduction, larger in
comparison with 12\% found in \citep{Villante:2013mba} and 
closer to 21\% found in  \citep{Bahcall:2000nu}.

\begin{figure*}\centering
\includegraphics[width=0.8\textwidth]{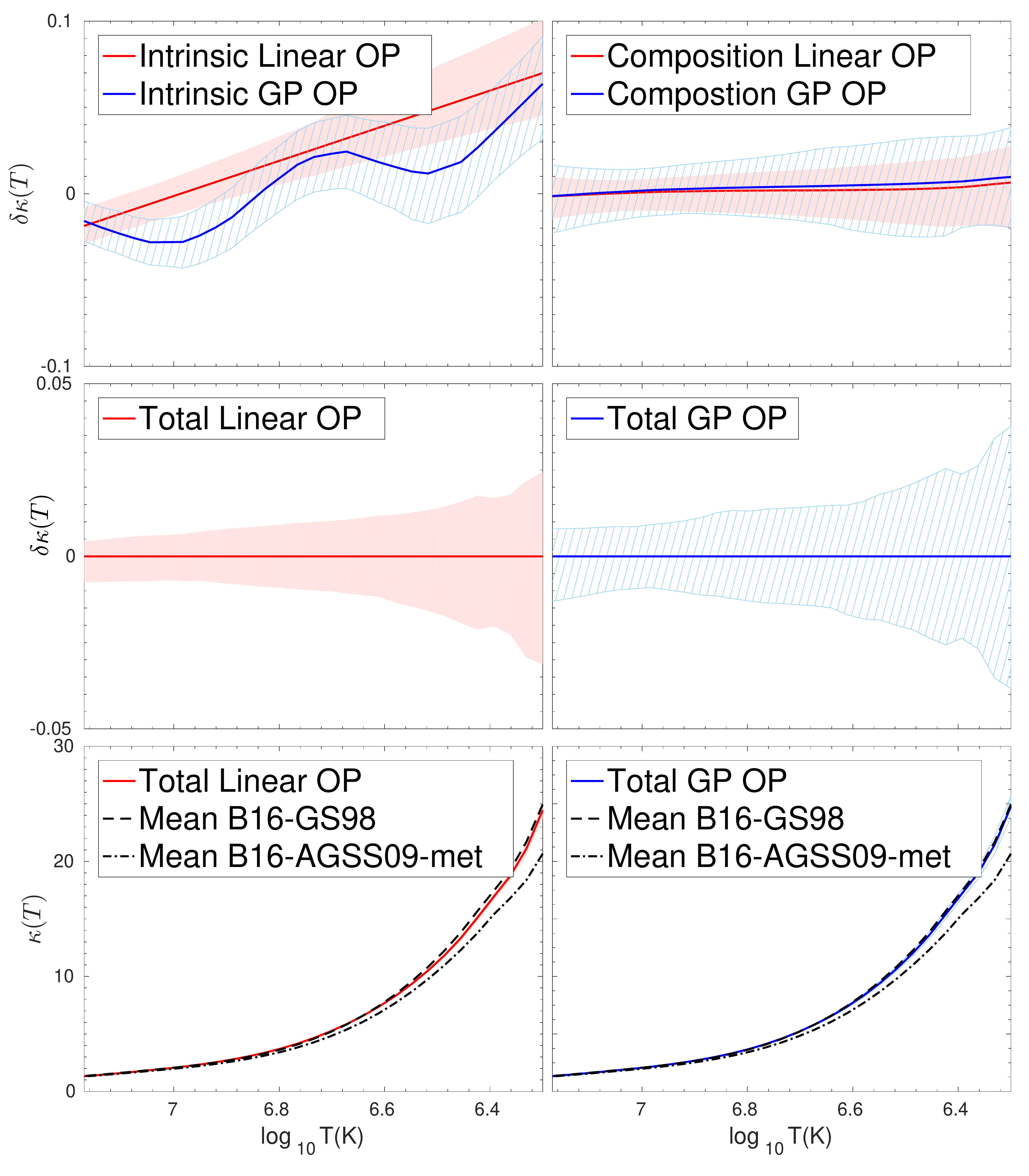}
\caption{
  Posterior distribution for the opacity profiles for the
  analysis with uniform priors for the abundances and the two choices of
  priors of the opacity uncertainties. See text for discussion.} 
\label{fig:opapostUNI}
\end{figure*}

The posterior distributions for the opacity profiles are shown in
Fig.~\ref{fig:opapostUNI}. In the upper left panel, we plot the
1$\sigma$ range of the intrinsic opacity change $\dk_I(T)$.
This is obtained from the posteriors of the parameters characterizing
this function, i.e. the parameters $a$ and $b$ for the linear opacity
parametrization given by Eq.~\eqref{eq:dkilin}, and the 11 values
$\delta \kappa_{Ii}=\delta \kappa_{I}(T_i)$ that sample the function
$\delta\kappa_I(r)$ (after marginalizing over the correlation length
$L$) for GP.
By construction, the intrinsic opacity change $\dk_I(T)$ is defined
with respect to a reference opacity calculation
$\overline{\kappa}(\rho,T,Y,Z_{i})$ that in our analysis include the
atomic opacities from OP \citep{Badnell:2004rz} complemented at low
temperatures by molecular opacities from \citep{Ferguson:2005pu}.
The fact that the posterior distribution of $\delta \kappa_{I}(T)$ is
not centered at zero (and, moreover, individuates a trend as a
function of $T$) indicates that there are features of the
observational data, namely the wiggle in the sound speed profile for
$0.3 < r/R_\odot < 0.6$, that cannot be optimally fitted by using the
reference opacities, even with the freedom of varying the solar input
parameters within their uncertainty ranges and the solar composition
in a large intervals considered in this paper, that accommodate both
AGSS09met and GS98 observational results.
The preference for a slight modification of the OP opacity is
consistent with what found in \citep{Villante:2013mba} where indeed it
was emphasized that the sound speed is better fitted by using the old
OPAL opacity tables.

As explained in Sec.~\ref{sec:opavscomp}, the quantity that is
directly constrained by observational data is the SSM opacity profile
$\kappa_{\rm SSM}(T)$, defined according to Eq.~\eqref{kappassm}, that
is affected by composition modifications (and solar model
recalibration) in addition to the effects of the intrinsic opacity 
change $\delta \kappa_{I}(T)$.
In the lower panels of Fig.~\ref{fig:opapostUNI}, we show the
posterior distributions for $\kappa_{\rm SSM}(T)$ for the linear
(left) and GP (right) description of opacity uncertainty.
The posterior distributions for $\kappa_{\rm SSM}(T)$ are compared
with the opacity profiles of B16-GS98 and B16-AGSS09met models.
We see that they are almost coincident with the the opacity profile of
B16-GS98 model, as it is expected by considering that the best fit CNO
and meteoritic elemental abundances, that drive the change in the
opacity, are close to GS98 determinations.
The optimal opacity profile is well defined by observational data, 
as it is seen in the central left (right) panels of Fig.~\ref{fig:opapostUNI} where we show 
the $1\sigma$ relative dispersion of $\kappa_{\rm SSM}(r)$ 
with respect to its mean posterior value.
The uncertainty for $\kappa_{\rm SSM}(r)$ is
somewhat larger for the GP opacity uncertainty description,
ranging from 0.8\% at the center to 4\% at the base of the convective
envelope, while for the linear  uncertainty parameterization it varies from
0.5\% to 2.5\%.

Finally, we note that the uncertainty in $\kappa_{\rm SSM}(r)$ is
smaller than that of the intrinsic opacity change.  In fact, $\delta
\kappa_{I}(r)$ is not directly constrained by the observational
properties of the Sun and its determination suffers from the
degeneracy with the composition opacity change $\delta \kappa_{\rm
Z}(T)$ that is quantified by Eq.~\eqref{deltakappa}. For completeness,
we report in the upper (right) panel of Fig.~\ref{fig:opapostUNI}, the
$1\sigma$ range for the composition opacity change $\delta
\kappa_{Z}(r)$, obtained from Eq.~\eqref{deltakappaZ} with $\delta
z_j$ being the variance of the posterior distributions of the
abundances in Fig.~\ref{fig:comppostUNI} defined relative to the mean
of those posteriors. Being defined with respect to the mean of the
posterior, the corresponding $\delta\kappa_Z$ are centered around zero.

\begin{figure}
\centering
\includegraphics[width=0.45\textwidth]{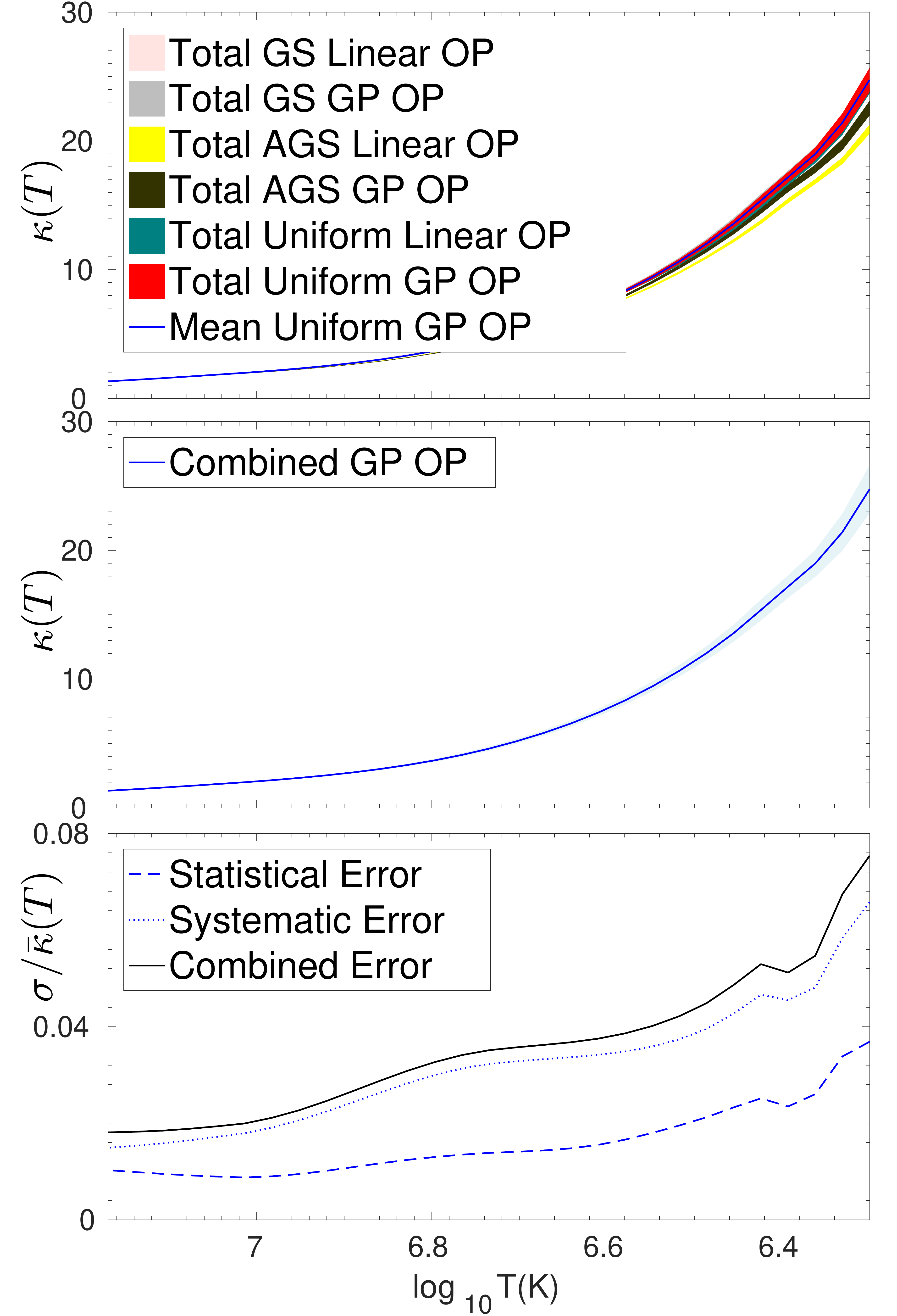}  
\caption{
{{\bf Top}: 1$\sigma$ ranges for the posteriors of the opacity
profiles corresponding to the six choice of priors for the composition
and intrinsic uncertainty variation used in this work.  {\bf Central}:
Posterior distribution for the opacity profiles for the analysis with
uniform priors for the abundances and the GP opacity uncertainty. The
panel shows the mean and 1$\sigma$ range of this distribution
combining both statistical and systematic uncertainties.  {\bf Lower}:
The panel shows its ``statistical'' 1$\sigma$ uncertainty defined as
the corresponding variance of the posterior (shown also as ``total''
in the central right panel in Fig.~\ref{fig:opapostUNI}) and its
``systematic'' uncertainty defined defined as the standard deviation
of the six profiles shown on the top window.}
\label{fig:finalopacity}}
\end{figure}

The result obtained with the uniform composition and with GP opacity uncertainty
prior represents our best estimate of the radiative opacity profile in the
solar interior. On the other hand, the profiles obtained with other choices
of priors, such us the uniform composition with linear opacity uncertainty,
or the four cases with B16-GS98 and B16-AGSSmet composition priors with either
choice of the opacity uncertainty prior presented in Sec.~\ref{sec:signif},
can serve as a  measure of the {\sl systematic} uncertainty in this estimate
that reflects dependence on the choice of priors. We show in the top panel
in Fig.~\ref{fig:finalopacity} the 1$\sigma$ range of the posteriors for
these six priors. From those we construct a systematic uncertainty in the
opacity, at each temperature, defined as the standard deviation of the six
reconstructed opacity profiles. The final opacity profile with both error
sources added in quadrature is shown in the central panel in Fig.~\ref{fig:finalopacity} and
it ranges from 2\% at the center to 7.5\% at the bottom of the convective
zone.

\begin{figure*}
\centering
\includegraphics[width=0.8\textwidth]{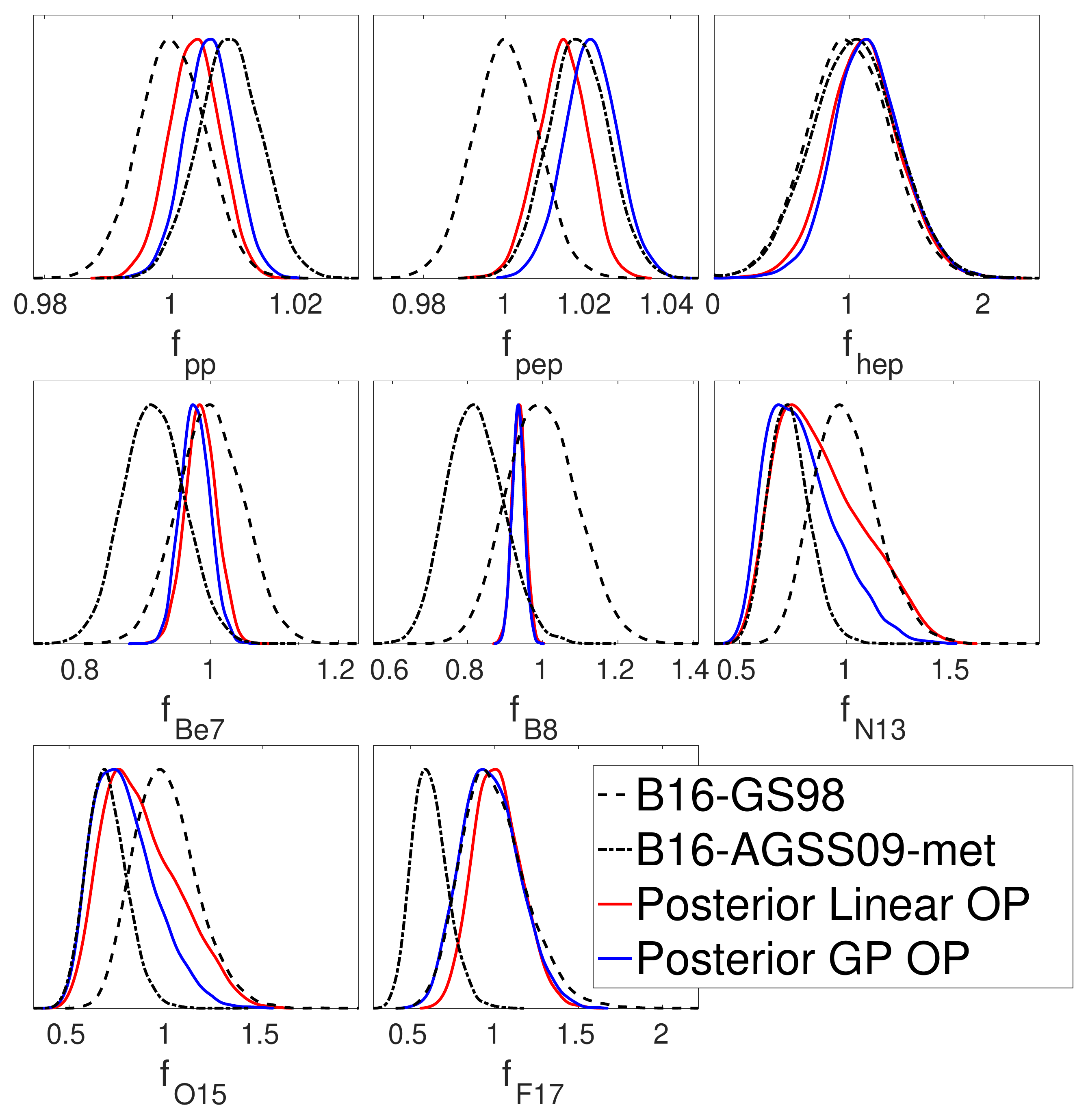}
\caption{Posterior distributions for the neutrino fluxes 
for the  analysis with uniform priors for the abundances and the
two choices of priors of the opacity uncertainties. For the sake of
comparison we show the corresponding priors for the
B16-GS98 and B16-AGSS09met models. The distributions are given in
arbitrary units and they have been normalized in such a way that the
maximum of all distributions lay at the same height.
The fluxes are shown normalized to the
B16-GS98 prediction $f_i=\phi_i/\phi_i^{\rm B16-GS98}$. }
\label{fig:fluxpostUNI}
\end{figure*}

Finally, for completeness, we show the resulting posterior distribution
for the neutrino fluxes in Fig.~\ref{fig:fluxpostUNI}. By construction
they constitute the predicted solar neutrino fluxes by models which
better describe both the helioseismic and neutrino data.  We denote them
as  helioseismic and neutrino data driven fluxes, B17-HNDD. We list
in Table~\ref{tab:fluxpostUNI}
their best values and 1$\sigma$ uncertainties
and in Eq.~\eqref{eq:postfluxrho} their correlations.
\setlength{\tabcolsep}{0.2em}
\begin{equation}
\rho=\left(\begin{tabular}{cccccccc}
1.00&0.80&0.03&-0.41&-0.02&-0.27&-0.27&0.12\\
0.80&1.00&0.06&-0.33&-0.05&-0.28&-0.29&0.01\\
0.03&0.06&1.00&-0.01&-0.01&-0.02&-0.02&0.01\\
-0.41&-0.33&-0.01&1.00&0.13&-0.03&-0.02&-0.03\\
-0.02&-0.05&-0.01&0.13&1.00&0.04&0.06&0.06\\
-0.27&-0.28&-0.02&-0.03&0.04&1.00&0.99&-0.14\\
-0.27&-0.29&-0.02&-0.02&0.06&0.99&1.00&-0.12\\
0.12&0.01&0.01&-0.03&0.06&-0.14&-0.12&1.00\\
\end{tabular}\right)
\label{eq:postfluxrho}
\end{equation}

\begin{table}
\centering
\setlength{\tabcolsep}{3pt}
\begin{tabular}{l|l}
  \multicolumn{2}{c}{B17-HNDD $\nu$-Fluxes}
  \\ \hline
 &   \\[-0.2cm]
$\phipp$ & $6.017\,(1^{+0.0033}_{-0.0041})$\\[+0.15cm]
$\phipep$  &$1.470\,(1\pm 0.0061)$\\[+0.15cm]
$\phihep$ & $9.04\,(1^{+0.22}_{-0.21})$\\[+0.15cm]
$\phibe$ &$4.79\,({1^{+0.027}_{-0.019}})$ \\[+0.15cm]
$\phib$ & $5.10\, (1\pm 0.018)$ \\[+0.15cm]
$\phin$ & $1.89\,(1^{+0.32}_{-0.14})$\\ [+0.15cm]
$\phio$ & $1.50\,(1^{+0.23}_{-0.20})$ \\[+0.15cm]
$\phif$ & $4.90\,(1^{+0.22}_{-0.18})$ \\[+0.15cm]\hline
\end{tabular}
\caption{Posterior solar neutrino fluxes for uniform-GP models. Units are:
  $10^{10}$ (pp), $10^9$ ($^7$Be), $10^8$ (pep, $^{13}$N, $^{15}$O),
  $10^6$ ($^8$B, $^{17}$ F) and
  $10^3$(hep) cm$^{−2}$s$^{−1}$.}
\label{tab:fluxpostUNI}
\end{table}
As expected, we find that for those neutrino fluxes which are at present most
precisely determined in solar neutrino experiments,
$^8$B and $^7$Be, the B17-HNDD flux  is very close to their experimental
value used to construct the  neutrino data part of the Likelihood function
(see last column in table~\ref{tab:ssmres}) but with a smaller uncertainty
because of the additional indirect constraints imposed by the helioseismic
data. Interestingly we find that with the inclusion of the helioseismic data
the precision of the predicted B17-HNDD CNO fluxes is only at most a factor
${\cal O}(2)$ weaker than those of the B16-GS98 or B16-AGSS09met composition
models.

\section{Summary}
\label{sec:summary}

In this work we have used Bayesian parameter inference and Gaussian process
for non-parametric functional reconstruction of the radial opacity
profile, with the goal of making an statistically consistent use of
the information from helioseismic and neutrino observations for solar modeling.
In particular to better determine the solar chemical composition and other
solar properties (as well as their uncertainties)  which are relevant to
the solar composition problem. 

In Secs.~\ref{sec:framework}  and ~\ref{sec:opa} we have presented
a brief summary of the statistical methodology followed and the application
of Gaussian process for functional reconstruction, and in particular to
the radial opacity profile parametrization. Sections ~\ref{sec:signif}
and ~\ref{sec:results} contain our results which we can be summarized
as follows:

\begin{itemize}

\item B16-GS98 vs B16-AGSS09met comparison. This improves over results
in \citep{Vinyoles:2016djt} because the linear parametrization of
opacities was not flexible enough. Now GP adds more flexibility to the
models so our results are now more general and much less dependent on
the choice of opacity tables. Helioseismic and neutrino data favors
the B16-GS98 model over B16-AGSS09met, but the more flexible modeling
of the opacity uncertainty allowed by the GP approach makes this
preference less marked.

\item Best composition. In our analysis all elements have uncorrelated
prior distributions.  Therefore, our results are more general than
those from previous works \citep{Villante:2013mba}.  When considering
individual elements, constraints are not very stringent on their
abundances.  This was expected. The best case is O, with a well
defined gaussian distribution with 1$\sigma$=0.07~dex, close to the
spectroscopic value.  When elements are grouped as CNO or meteoritic,
the posterior distributions of these groups are well peaked with
uncertainties in the linear(GP) analysis of 0.025(0.045)~dex and
0.01(0.015)~dex respectively, comparable to those obtained from
spectroscopic measurements.  Due to our adoption of a flat prior for
elemental abundances and our introduction of the GP approach for
modeling opacity uncertainties, our results are quite general, with as
little dependence on modeling assumptions as possible (e.g. the bounds in in
\citealt{Villante:2013mba} are obtained in the assumption that the 
difference OP-OPAL is the measure of the intrinsic opacity uncertainty).

\item Non-composition input parameters. The posterior distributions of
these parameters have also been determined and are the most general
results available to date.  $S_{11}$ varies at the 1$\sigma$ level
(1\% with respect to \citep{Marcucci:2013tda} when compared to
\citep{Acharya:2016kfl}. This is not a large difference, but further
work on this important rate might be worth. Our best estimate of the
rate of microscopic diffusion is also lower, by about 2$\sigma$, than
the standard rate used in solar models. This is qualitatively
expected, but the 30\% reduction is quantitatively larger than
previous estimates that suggested reductions in the range of 15-20\%
\citep{Delahaye:2005ed,Villante:2013mba}

\item Opacity reconstruction. This is the most important result of our
work.  We have been able to reconstruct the solar opacity profile in a
data driven way, i.e. without strong assumptions on the solar
composition or the underlying opacity tables. 
Considering uncertainties due to the solar data alone, the
opacity uncertainty is about 4\% at the base of the convective zone
and less than 1\% at the solar core.  Different sets of priors help us
quantify a systematic uncertainty in this estimate. From a broad range
of assumptions, our more conservative estimate of the total opacity
uncertainty (data + priors) is 7.5\% at the base of the convective
envelope and 1.8\% at the solar core.
  
\item Neutrino fluxes. We have obtained the posterior distributions of
solar $\nu$-fluxes based on the uniform prior distribution of solar
abundances and GP treatment of opacity uncertainties. These fluxes
represent the best data driven reconstruction of the expected solar
models $\nu$-fluxes. For the well measured $^8$B and $^7$Be fluxes,
the final uncertainties reflect experimental uncertainties. For CN
fluxes, the predicted values are approximately only a factor of 2
larger than in the B16 SSMs ($\sim$\,20-25\%). This is remarkable
because their uncertainty is dominated in our analysis by the C+N
abundance, that has a much larger prior range of variation.

\end{itemize}

\section*{acknowledgments} We thank C. Pe\~na-Garay and the SOM group
(University of Valencia) for their important collaboration in the
early stages of this work in particularly for providing us with the
computational power to compute the x10000 SSMs calculations.  We also
thank Michele Maltoni for discussions.  We are specially indebted to
Johannes Bergstrom for introducing us to the use of Gaussian Processes
and \MN\ and personally initiating this project.  This work is
supported by USA-NSF grant PHY-1620628, by EU Networks FP10 ITN
ELUSIVES (H2020-MSCA-ITN-2015-674896) and INVISIBLES-PLUS
(H2020-MSCA-RISE-2015-690575), by MINECO grants FPA2016-76005-C2-1-P
and ESP2015-66134-R, and by Maria de Maetzu program grant 
MDM-2014-0367 of ICCUB.

\bibliographystyle{mnras}
\bibliography{references}

\appendix
\section{Data included in the analysis}
\label{sec:appendix}
We construct the likelihood function with data from helioseismology
and neutrino oscillation experiments. 
In particular we include the two 
helioseismic quantities widely used in assessing the quality of
SSMs: the surface helium abundance $\ysur$ and the location of the
bottom of the convective envelope $\rcz$.
In Table~\ref{tab:ssmres} we
include the experimentally determined value for those two quantities.
For illustration we also show the mean and variation of their expected values
in the B16 SSM's (which however are not directly use in building
the corresponding likelihood). In building the
corresponding likelihood function we assume the experimental errors to be
totally uncorrelated.
\begin{table}
\setlength{\tabcolsep}{2.5pt}
\begin{tabular}{l|ccc}
Qnt. & B16-GS98 & B16-AGSS09met & Solar \\
\hline
$\ysur$& $0.2426 \pm 0.0059 $&$0.2317 \pm 0.0059 $& $0.2485 \pm 0.0035$ \\
$\rcz/\rsun$& $0.7116 \pm 0.0048$ & $0.7223  \pm 0.0053$ & $0.713 \pm 0.001 $\\
\hline
$\phipp$ & $5.98(1 \pm 0.006)$ & $6.03(1 \pm 0.005) $&$5.971^{(1+0.006)}_{(1-0.005)}$\\
$\phipep$  &$ 1.44(1 \pm 0.01) $&$1.46(1 \pm 0.009) $&$1.448 (1\pm 0.009)$\\
$\phihep$ & $7.98(1 \pm 0.30) $&$8.25(1 \pm 0.30) $&$19^{(1+0.63)}_{(1-0.47)}$\\
$\phibe$ &$ 4.93(1 \pm 0.06)$ &$4.50(1 \pm 0.06) $&$4.80^{(1+0.050)}_{(1-0.046)}$\\
$\phib$ & $5.46(1 \pm 0.12)$ &$4.50(1  \pm 0.12) $&$5.16^{(1+0.025)}_{(1-0.017)}$\\
$\phin$ & $2.78(1 \pm 0.15)$ &$2.04(1  \pm  0.14) $&$\le 13.7$\\
$\phio$ & $2.05(1 \pm 0.17)$ &$1.44(1 \pm 0.16) $&$\le 2.8$\\
$\phif$ & $5.29(1 \pm 0.20)$ &$3.26(1 \pm 0.18) $&$\le 85$\\ \hline
\end{tabular}
\caption{Main characteristics for the different SSMs with the
  correspondent model errors and the values for the observational
  values and their error. For the fluxes units are:
  $10^{10}$ (pp), $10^9$ ($^7$Be), $10^8$ (pep, $^{13}$N, $^{15}$O),
  $10^6$ ($^8$B, $^{17}$ F) and
  $10^3$(hep) cm$^{−2}$s$^{−1}$. For the fluxes the last column  ``Solar''
  corresponds to the
  values obtained from direct fit to the solar neutrino data in
   \citet{Bergstrom:2016cbh}.}
  \label{tab:ssmres}
\end{table}

\begin{figure}
\includegraphics[width=0.48\textwidth]{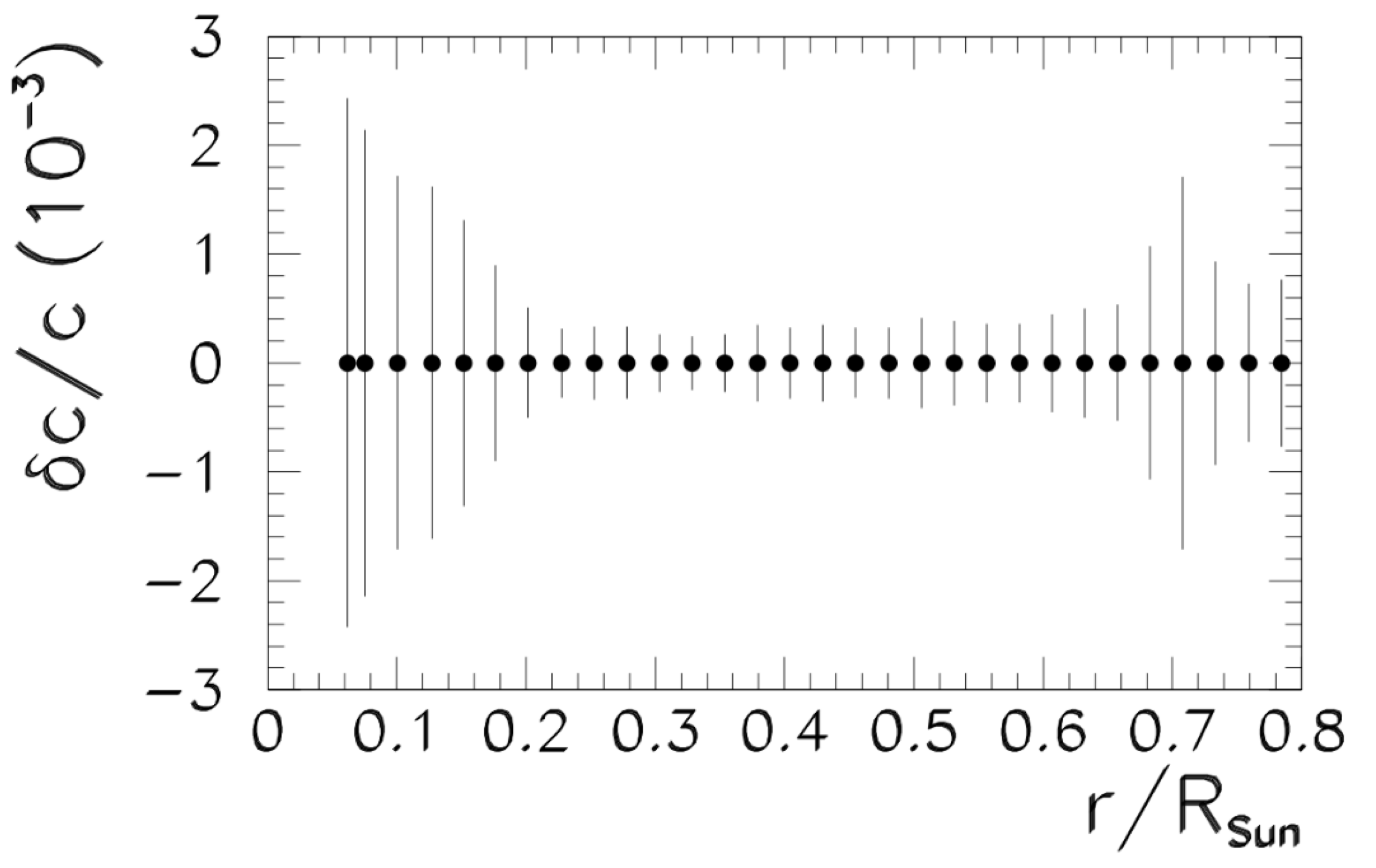}
\caption{
  Fractional sound speed data
  $\delta c/c=(c_\odot-\bar c^{\rm mod})/\bar c^{\rm mod}$
  used in our analysis.
\label{fig:cs}}
\end{figure}

We also include the fractional sound speed differences
in 30 points along the solar radius determined by performing sound speed
inversions as described in  \citep{Basu:2009mi}.
In \citet{Vinyoles:2016djt} we give a detailed summary of the sources
of uncertainties for the sound speed profile. These  ``experimental'' uncertainties
are conservatively assumed to be uncorrelated. For completeness we plot 
in Fig.~\ref{fig:cs} the fractional sound speed
differences used in our statistical analysis which, by definition, have zero 
central values.

Finally we include the results from oscillation experiments in the form
of the likelihood of the global analysis of neutrino
oscillation data used and described in \citet{Bergstrom:2016cbh} in terms of
3-$\nu$ oscillations with arbitrary normalization of each of the components
of the solar flux. For the sake of illustration we list in
the last column in Table~\ref{tab:ssmres} the central values and errors of
the solar flux normalizations extracted in that analysis.
Effectively the effect of the inclusion of the neutrino oscillation data
can be understood in terms of a reduced gaussian likelihood constructed
with these extracted eight  solar fluxes and uncertainties and with the 
correlation matrix:
\setlength{\tabcolsep}{0.3em}
\begin{equation}
\rho= 
 \left(\begin{tabular}{cccccccc}
  1&0.99&-0.05&0.08&-0.14&-0.20&-0.19&-0.11\\
 0.99&1&-0.05&0.08&-0.14&-0.20&-0.19&-0.11\\
 -0.05&-0.05&1&-0.08&0.10&-0.01&-0.00&-0.00\\
 0.08&0.08&-0.08&1&-0.17&-0.31&0.09&0.10\\
 -0.14&-0.14&0.10&-0.17&1&-0.02&-0.03&-0.01\\
 -0.20&-0.20&-0.01&-0.31&-0.02&0&0.18&0.09\\
 -0.19&-0.19&-0.00&0.09&-0.03&0.18&1&0.36\\
 -0.11&-0.11&-0.00&0.10&-0.01&0.09&0.36&1\\
 \end{tabular}\right)
 \label{eq:rhonu}
 \end{equation}

\section{Gaussian Process for Function Reconstruction}
\label{sec:GP}

Gaussian process (GP) is a non-parametric regression method widely used
in statistics and machine learning to reconstruct a function which
best describes some data  without assuming a parametrization of
the function (see e.g. \citealt{Murphy12,Rasmussen06,MacKay03} or
the Gaussian Process webpage\footnote{\url{http://www.gaussianprocess.org/}} for details).
It is used for example in data analysis in cosmology to reconstruct some of
the evolution dependent properties (like the dark energy equation of state; 
 \citealp{Holsclaw:2010nb,Holsclaw:2010sk,Seikel:2012uu}). 
 \citet{Seikel:2012uu} contains a pedagogical description of the process
which we  briefly sketch here. 

The starting assumption is that the value of the function $f$
evaluated at a point $x$ is a Gaussian random variable of mean $\mu(x)$ and
variance ${\rm Var}(x)$. As the values of the function in two points
$x$ and $x'$ are not independent, in general one can define a covariant
function ${\rm cov}(f(x),f(x'))\equiv C(x,x')$. The assumed ``prior'' covariance
function is arbitrary although the obvious hypothesis is that it depends
only on the distance between the points. For example, a common choice is
a square exponential
\begin{equation}
C_P(x,x')=\sigma_f^2\,\mathrm{exp}\left(-\frac{(x-x')^2}{2\lambda^2}\right),
\label{eq:covGP}
\end{equation}
which depends on the parameters $\sigma_f$ and $\lambda$, often referred to as
``hyperparameters'' as they do not specify the form of the function
but give a measure of its characteristic variations. $\lambda$ can be seen as
the characteristic length over which the function changes significantly
while $\sigma_f$ is its range of variation at each point.

The procedure aims at determining the posterior mean and variance value 
of the function at some predetermined points, $f_i=f(x_i)$, $i=1,N$. 
This is, to determine $\mu_i=\mu(x_i)$ and $C_{ij}=C(x_i,x_j)$ starting
from some prior mean function $\mu_p(x)$ and the chosen prior for the
covariant function. 
It does so by finding the optimum values of $\sigma_f$ and $\lambda$
(or  marginalizing over them) by confronting them with the data.

In the simplest case the data to be described corresponds to the value
of the function at specific points $\tilde x_a$, $y_a=f(\tilde x_a)$
with $a=1$ to $\tilde N$, known with some
uncertainties $\sigma_a$ (or what is the same with some experimental
covariance $\tilde C_{ab}$). In this case it can be shown that the likelihood
for the hyperparameters takes the form
\begin{eqnarray}
&& -2 \ln \lhood(\sigma_f,L)= \\
&&  \sum_{ab=1}^{\tilde N}\left\{(y_a-\mu_a)(C_t)^{-1}_{ab}(y_b-\mu_b)+
  \ln (C_t)_{ab} \right\}  +\, \rm{const} \nonumber
\end{eqnarray}
where $\mu_a=\mu_p(x_a)$ and $C_t = C_P + \tilde C$.  
The posterior mean and covariance for the function at the specific points are 
\begin{eqnarray}
\bar f_i=\mu_p(x_i)+ \sum_{ab=1}^{\tilde N}
  {C_P}_{ia}(C_t)^{-1}_{ab}(y_b-\mu_b) \\
C_{ij}={C_P}_{ij}-\sum_{ab=1}^{\tilde N} {C_P}_{ia}(C_t)^{-1}_{ab} C_{bj}\, .    
\end{eqnarray}

For the problem at hand, the data -- neutrino fluxes, helioseismic data,
and sound speeds -- are functions of the opacity function that we want
to determine (not some values of it) so the procedure to use Gaussian Process
has to be adapted as described in Sec.~\ref{sec:GPop}

\bsp	
\label{lastpage}
\end{document}